\documentclass{jfm}
\usepackage{amsfonts}
\usepackage{amsmath}
\usepackage{amssymb}
\usepackage{graphicx}
\usepackage{epstopdf, epsfig}
\usepackage[round,super]{natbib}
\usepackage[pdfstartview=FitH, CJKbookmarks=true,
bookmarksnumbered=true, bookmarksopen=true,
colorlinks,
pdfborder=001, linkcolor=green,
anchorcolor=blue, citecolor=blue
]{hyperref}
\usepackage{filecontents}

\newcommand{\comment}[1]{}

\shorttitle{Modulation-resonance mechanism for surface
waves}
\shortauthor{S. W. Jiang, G. Kova\v{c}i\v{c}, D. Zhou and D. Cai}
\affiliation{\aff{1}Department of Mathematics, the Pennsylvania State University, University Park, PA 16802-6400, USA
\aff{2}Mathematical Sciences Department, Rensselaer Polytechnic Institute,
110 8th Street, Troy, New York 12180, USA
\aff{3}School of Mathematical Sciences, MOE-LSC, and Institute of Natural Sciences,
Shanghai Jiao Tong University, Shanghai 200240, China
\aff{4}Courant Institute of Mathematical Sciences and Center for Neural
Science, New York University, New York, New York 10012, USA
\aff{5}NYUAD Institute, New York University Abu Dhabi, P.O. Box 129188, Abu
Dhabi, United Arab Emirates}

\begin{document}

\title{Modulation-resonance mechanism for \\
surface waves in a two-layer fluid system}
\author{Shixiao W. Jiang\aff{1} \corresp{\email{suj235@psu.edu}}, Gregor Kova%
\v{c}i\v{c}\aff{2} \corresp{\email{kovacg@rpi.edu}}, Douglas Zhou\aff{3} %
\corresp{\email{zdz@sjtu.edu.cn}} \and David Cai\aff{3,4,5} }
\maketitle

\begin{abstract}
We propose a Boussinesq-type model to study the surface/interfacial wave
manifestation of an underlying, slowly-varying, long-wavelength, baroclinic
flow in a two-layer, density-stratified system. The results of our model
show numerically that, under strong nonlinearity, surface waves, with their
typical wavenumber being the resonant $k_{\mathrm{res}}$, can be generated
locally at the leading edge of the underlying slowly-varying,
long-wavelength baroclinic flow. Here, the resonant $k_{\mathrm{res}}$
satisfies the class 3 triad resonance condition among two short-mode waves
and one long-mode wave in which all waves propagate in the same
direction. Moreover, when the slope of the baroclinic flow is
sufficiently small, only one spatially-localized large-amplitude surface
wave packet can be generated at the leading edge. This localized surface
wave packet becomes high in amplitude and large in group velocity after the
interaction with its surrounding waves. {These results are qualitatively
consistent with various experimental observations including resonant surface
waves at the leading edge of an internal wave.} Subsequently, we propose a
mechanism, referred to as the modulation-resonance mechanism, underlying
these surface phenomena, based on our numerical simulations. The proposed
modulation-resonance mechanism combines the linear modulation (ray-based)
theory for the spatiotemporal asymmetric behavior of surface waves and
the nonlinear class 3 triad resonance theory for the energy focusing of
surface waves around the resonant wavenumber $k_{\mathrm{res}}$ in Fourier
space.
\end{abstract}


\section{Introduction}

Large-amplitude internal waves with long wavelengths that propagate over
long distances are a universal phenomenon in density-stratified coastal
ocean regions, and they play an important role in transferring momentum, heat,
and energy in those regions
\citep[][and references
therein]{Alford2015Nature,Duda1998DTIC}. Internal waves are not directly
visible to observers; their presence can be inferred using the scattering of
radar illumination from their accompanying surface waves (SWs). In recent decades,
observation data for the surface signature of internal waves have been
recorded using Synthetic Aperture Radar (SAR) in many coastal seas
worldwide. However, the mechanism underlying the surface signature of
internal waves has not been completely understood.

{In many field observations}, there exists a significant disparity between the
typical scales of internal waves and their accompanying SWs: the typical
wavelength of internal waves is several kilometres and the typical amplitude
is about $100$ metres \citep{Perry1965JGR,Duda2004IEEEJOE}, while on the other
hand, the accompanying SWs are between a decimetre and a metre in wavelength %
\citep{Gargettt1972JFM,Bakhanov2002JGR,Hwung2009JFM}. A basic surface
manifestation of internal waves consists of wide bands of smooth surface
(slicks), alternating with narrow bands of rough surface, possibly
containing breaking waves (roughness) %
\citep[][]{Gargettt1972JFM,Alpers1985Nature}. 
The typical features of the internal-wave surface signature reported in
various field observations can be summarized as follows:

\hypertarget{ASYM_MODU}
\noindent (i) \textit{Surface roughness at the leading edge of internal waves%
}\label{asym_modu}: The surface ripples were found to be located at the
leading edge of an internal wave %
\citep[][]{Osborne1980Sci,Alpers1985Nature,Gasparovic1988JGR,Bakhanov2002JGR,Moore2007MMS}%
. In the narrow rough regions, due to the tendency of SWs to shorten and
steepen, they are more likely to break and then leave behind calm water
after the internal-wave passage %
\citep{Phillips1966Dynamics,Gasparovic1988JGR,Craig2012JFM}.

\hypertarget{CLASS_RESON}
\noindent (ii) \textit{Resonant excitations}\label{class3_reson}: The rough
region travels along with the internal wave; that is, the SW group velocity
is close to the internal-wave phase velocity %
\citep{Osborne1980Sci,Kropfli1999JGR,Hwung2009JFM,Craig2012JFM}.

{\noindent To understand the above two surface phenomena, we numerically
study the surface-wave and interfacial-wave (IW) manifestation in a
two-layer, density-stratified system using a weakly nonlinear  Boussinesq-type model.
In this work, we focus on qualitatively capturing the above two phenomena of
the surface signature reflecting an underlying IW. Subsequently, we propose a
mechanism, referred to as the modulation-resonance mechanism,
inferred from our simulation to understand the dynamical behavior of these SWs.}

{A large volume of literature is dedicated to two-layer fluid systems
modeling the interaction between interfacial and surface waves.%
} These works can be broadly grouped into two classes. The first class
consists of the literature addressing the linear modulation (or ray-based)
theory. These ray-based studies take a statistical viewpoint of SWs
modulated by a near-surface current induced by IWs, and invoke
phase-averaged models based on a wave-balance equation and ray-based theory %
\citep[e.g.][]{Gargettt1972JFM,Lewis1974JFM,Caponi1988DTIC,Bakhanov2002JGR,Apel2007JASA,Kodaira2016JFM}%
. The second class consists of the literature addressing the nonlinear
resonance theory, in particular, the class 3 triad resonance theory \citep{Lewis1974JFM,Hashizume1980JPSJp,Alam2012JFM,Craig2012JFM,Tanaka2015JFM}.
These studies are based on the possibility of a resonant
interaction between multiple modes which coexist in a two-layer
density-stratified fluid system %
\citep{Phillips1974ARFM,Lewis1974JFM,Benney1977SAM}. Many of the theoretical
studies are dedicated to the effective model describing resonant
coupling between interfacial and surface waves, which is derived from
two-layer Euler equations using the triad resonance condition via a
multi-scale analysis %
\citep{Lewis1974JFM,Kawahara1975JPSJ,Hashizume1980JPSJp,Funakoshi1983JPSJp,Sepulveda1987PhysFld,Lee2007JKrPS,Hwung2009JFM,Craig2012JFM}%
.
In contrast, only a few of numerical-simulation studies
concern the triad resonant phenomena %
\citep{Funakoshi1983JPSJp,Hwung2009JFM,Craig2012JFM,Tanaka2015JFM}, such as the inverse
energy cascade phenomenon for SWs \citep{Tanaka2015JFM}.
{However,
it is still challenging to simultaneously capture the above
surface phenomena
\hyperlink{ASYM_MODU}{(i)} and \hyperlink{CLASS_RESON}{(ii)}
in these previous theories and numerics.}

In this work, we focus on the qualitative behavior of SWs when there
is an underlying baroclinic flow, and use numerical simulations
to mimic and understand the field observations of
\hyperlink{ASYM_MODU}{(i)}
surface ripples located at the leading edge and
\hyperlink{CLASS_RESON}{(ii)}
resonant excitations of these surface ripples.
Our two-layer model predicts that,
in the strongly nonlinear
regime, a finite number of localized SW packets can be generated at the
leading edge of the background baroclinic flow (BBF) with their group velocity
being close to the phase velocity of the BBF.
To examine the robustness of our model results, we investigate
the behavior of these SWs for various underlying BBFs in a
parametric study. In particular, if the slope of the
BBF is sufficiently small, only one localized SW packet or very few localized SW
packets can be generated at the leading edge,  and they become high in amplitude and large in
group velocity after interacting with the surrounding waves.

Importantly, we propose the modulation-resonance
mechanism to understand the above surface phenomena.
The proposed mechanism combines two different mechanisms:
[\textit{modulation mechanism}] First, based on the linear modulation (ray-based) theory,
SW packets  propagate quickly towards the leading edge of the BBF and a sink of SW packets
is thereafter formed at the leading edge. [\textit{resonance mechanism}]
Subsequently, based on the class 3 triad resonance, the spectrum of SWs
can eventually be concentrated near the resonant wavenumber in the wavenumber $k$ space,
and simultaneously these resonant SWs become high in amplitude at the leading edge
in the spatial domain.
The modulation-resonance
mechanism may be extended to general density-stratified fluid systems
in both shallow-water and deep-water configurations.







The paper is organized as follows. In \S\ \ref{sec:one_dim_isw}, we
formulate the density-stratified-fluid problem, and provide a
weakly-nonlinear Boussinesq-type model describing the coupling between IWs and free SWs. In
\S\ \ref{sec:review_lin_non} and \S\ \ref{sec:strong_nonlinear},
we numerically investigate the qualitatively-consistent IW surface signatures
\hyperlink{ASYM_MODU}{(i)} and \hyperlink{CLASS_RESON}{(ii)}
using our model,
and further
propose the modulation-resonance mechanism underlying these surface
manifestations. In \S\ \ref{sec:conclusion}, we present conclusions and
discussion.
We relegate some technical details  to the appendices.

\section{Formulation of the two-layer fluid system}

\label{sec:one_dim_isw}

\subsection{The two-layer weakly nonlinear (TWN) model}

\begin{figure}
\centering 
\hspace*{-2.0mm} \includegraphics[width=0.70\textwidth,height=0.33%
\textheight]{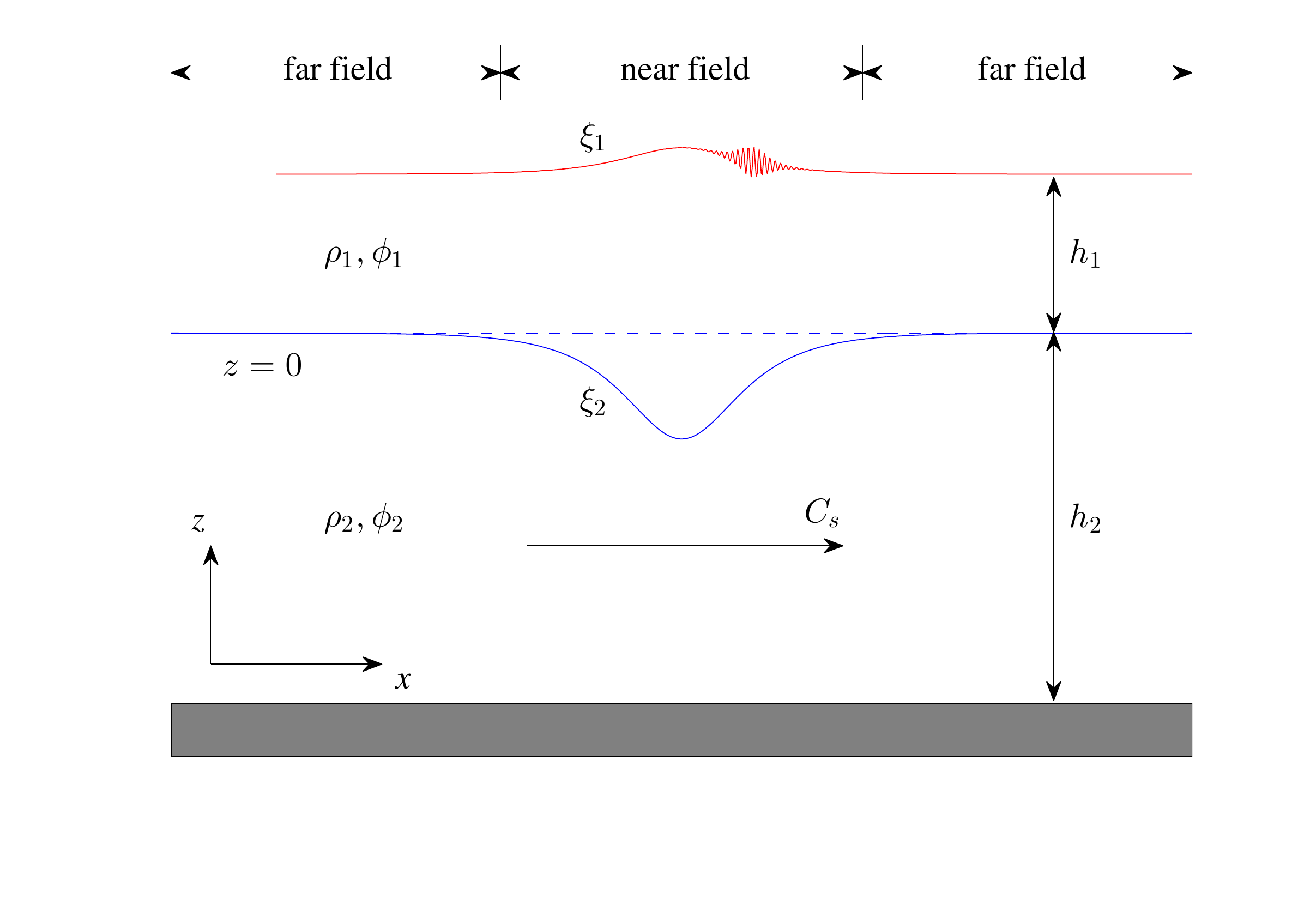}
\caption{(Colour online) Sketch of the two-layer fluid system. The surface
ripples are located at the leading edge of the IW. Near field
corresponds to the region near the central IW and far field
corresponds to the region far from the central IW.}
\label{Fig:sketch_two_layer}
\end{figure}

We begin with Euler equations for two immiscible layers of potential fluids
with unequal densities. The fluids in the two layers are assumed to be
inviscid, irrotational, and incompressible, with unequal densities, $\rho
_{1}$ in the upper layer and $\rho _{2}$ ($>\rho _{1}$ for the
stable case) in the lower layer. The horizontal and vertical coordinates
are denoted by $x$ and $z$, respectively. The interface and the overlaying
free-surface displacements are denoted, respectively, by $\xi _{2}$ and $\xi
_{1}$ (see figure \ref{Fig:sketch_two_layer}). The pressure exerted at the
surface is denoted by $P_{1}$.
Then, the governing equations for the two-layer Euler system are given as
follows:
\begin{equation}
\phi _{1xx}+\phi _{1zz}=0,\text{ \ \ }\xi _{2}<z<h_{1}+\xi _{1},
\label{Eqn:1.Laplace}
\end{equation}%
\begin{equation}
\phi _{2xx}+\phi _{2zz}=0,\text{ \ \ }-h_{2}<z<\xi _{2},
\label{Eqn:1.lowLaplace}
\end{equation}%
\begin{equation}
\xi _{1t}+\phi _{1x}\xi _{1x}=\phi _{1z},\text{ \ \ at }z=h_{1}+\xi _{1},
\label{Eqn:1.kin_sf}
\end{equation}%
\begin{equation}
\phi _{1t}+\frac{1}{2}\left( \phi _{1x}^{2}+\phi _{1z}^{2}\right) +g\xi
_{1}+P_{1}/\rho _{1}=0,\text{ \ \ at }z=h_{1}+\xi _{1},  \label{Eqn:1.dyn_sf}
\end{equation}%
\begin{equation}
\xi _{2t}+\phi _{1x}\xi _{2x}=\phi _{1z},\text{ \ \ at }z=\xi _{2},
\label{Eqn:1.kin_if1}
\end{equation}%
\begin{equation}
\xi _{2t}+\phi _{2x}\xi _{2x}=\phi _{2z},\text{ \ \ at }z=\xi _{2},
\label{Eqn:1.kin_if2}
\end{equation}%
\begin{align}
& \rho _{1}\left( \phi _{1t}+\frac{1}{2}\left( \phi _{1x}^{2}+\phi
_{1z}^{2}\right) +g\xi _{2}\right)  \label{Eqn:1.dyn_if} \\
& =\rho _{2}\left( \phi _{2t}+\frac{1}{2}\left( \phi _{2x}^{2}+\phi
_{2z}^{2}\right) +g\xi _{2}\right) ,\text{ \ \ at }z=\xi _{2},  \notag
\end{align}%
\begin{equation}
\phi _{2z}=0,\text{ \ \ at }z=-h_{2},  \label{Eqn:1.kin_bm}
\end{equation}%
where $\phi _{1}$ ($\phi _{2}$) is the velocity potential of the upper
(lower) fluid layer, $h_{1}$ ($h_{2}$) is the undisturbed thickness of the
upper (lower) fluid layer, respectively, and $g$ is the gravitational
acceleration.

For the small-amplitude approximation, we assume that the characteristic
amplitude, $a$, of the IWs and SWs is much smaller than the thickness of the
upper fluid layer,
\begin{equation}
a/h_{1}=\alpha \ll 1.  \label{Eqn:1.assm_alpha}
\end{equation}%
For the long-wave approximation, we assume that the thickness of each fluid
layer is much smaller than the characteristic wavelength, $l$, of the IWs
and SWs,%
\begin{equation}
h_{1}^{2}/l^{2}=\beta = O(\alpha) \ll 1,\text{\ \ \ }h_{2}=O(h_{1}),
\label{Eqn:1.assm_beta}
\end{equation}%
The two small parameters, $\alpha $ and $\beta $, control the nonlinear and
dispersive effects, respectively.
Based on the scaling (\ref{Eqn:1.assm_alpha})-(\ref{Eqn:1.assm_beta}), we nondimensionalize all the physical variables in the Euler equations (\ref{Eqn:1.Laplace})-(\ref{Eqn:1.kin_bm}).
Next, we apply asymptotic analysis to these nondimensionalized equations, similar to that applied in the derivation of the KdV equation \citep{Whitham1974Linear}.
Retaining the terms in the first power of $\alpha$ and $\beta$, \hypertarget{M1}{we} obtain the set of
equations in the dimensional form for the variables $(\xi _{1},\overline{u}%
_{1},\xi _{2},\overline{u}_{2})$ [to which we refer as the two-layer weakly-nonlinear
model or TWN model], \label{M1_ref}
\begin{equation}
\eta _{1t}+(\eta _{1}\overline{u}_{1})_{x}=0,\text{ \ \ }\eta _{1}=h_{1}+\xi
_{1}-\xi _{2},  \label{Eqn:1.upkin}
\end{equation}%
\begin{equation}
\eta _{2t}+(\eta _{2}\overline{u}_{2})_{x}=0,\text{ \ \ }\eta _{2}=h_{2}+\xi
_{2},  \label{Eqn:1.dnkin}
\end{equation}%
\begin{equation}
\overline{u}_{1t}+\overline{u}_{1}\overline{u}_{1x}+g\xi _{1x}-\frac{1}{3}%
h_{1}^{2}\overline{u}_{1xxt}+\frac{1}{2}h_{1}\xi _{2xtt}+P_{1x}/\rho _{1}=0,
\label{Eqn:1.updyn}
\end{equation}%
\begin{align}
& \overline{u}_{2t}+\overline{u}_{2}\overline{u}_{2x}+g\xi _{2x}+\rho
_{r}g\eta _{1x}  \label{Eqn:1.dndyn} \\
& -\frac{1}{2}\rho _{r}h_{1}^{2}\overline{u}_{1xxt}+\rho _{r}h_{1}\xi
_{2xtt}-\frac{1}{3}h_{2}^{2}\overline{u}_{2xxt}+P_{1x}/\rho _{2}=0,  \notag
\end{align}%
where $\rho _{r}$ is the density ratio $\rho _{1}/\rho _{2}$, and
\begin{equation*}
\overline{u}_{1}(x,t)=\dfrac{1}{\eta _{1}}\int_{\xi _{2}}^{h_{1}+\xi
_{1}}\phi _{1x}(x,z,t)dz,\text{ \ \ }\overline{u}_{2}(x,t)=\dfrac{1}{\eta
_{2}}\int_{-h_{2}}^{\xi _{2}}\phi _{2x}(x,z,t)dz,
\end{equation*}%
are the depth-averaged horizontal velocities of the upper and lower fluid
layers, respectively. This weakly-nonlinear model can also be obtained via a
direct reduction from a fully nonlinear model given in \citet{Choi1996JFM}, %
\citet{Barros2007SAM}, and \citet{Barros2009SAM}. We refer to the fully nonlinear model
as the MCC model. We provide the dimensionless
form of the TWN model in appendix \ref{subsec:limitation}. Next, we study the
basic properties of the TWN model, including
the dispersion relation and the class 3 triad resonance condition.

\begin{figure}
\centering 
\hspace*{-10.0mm} %
\includegraphics[scale=1.0]{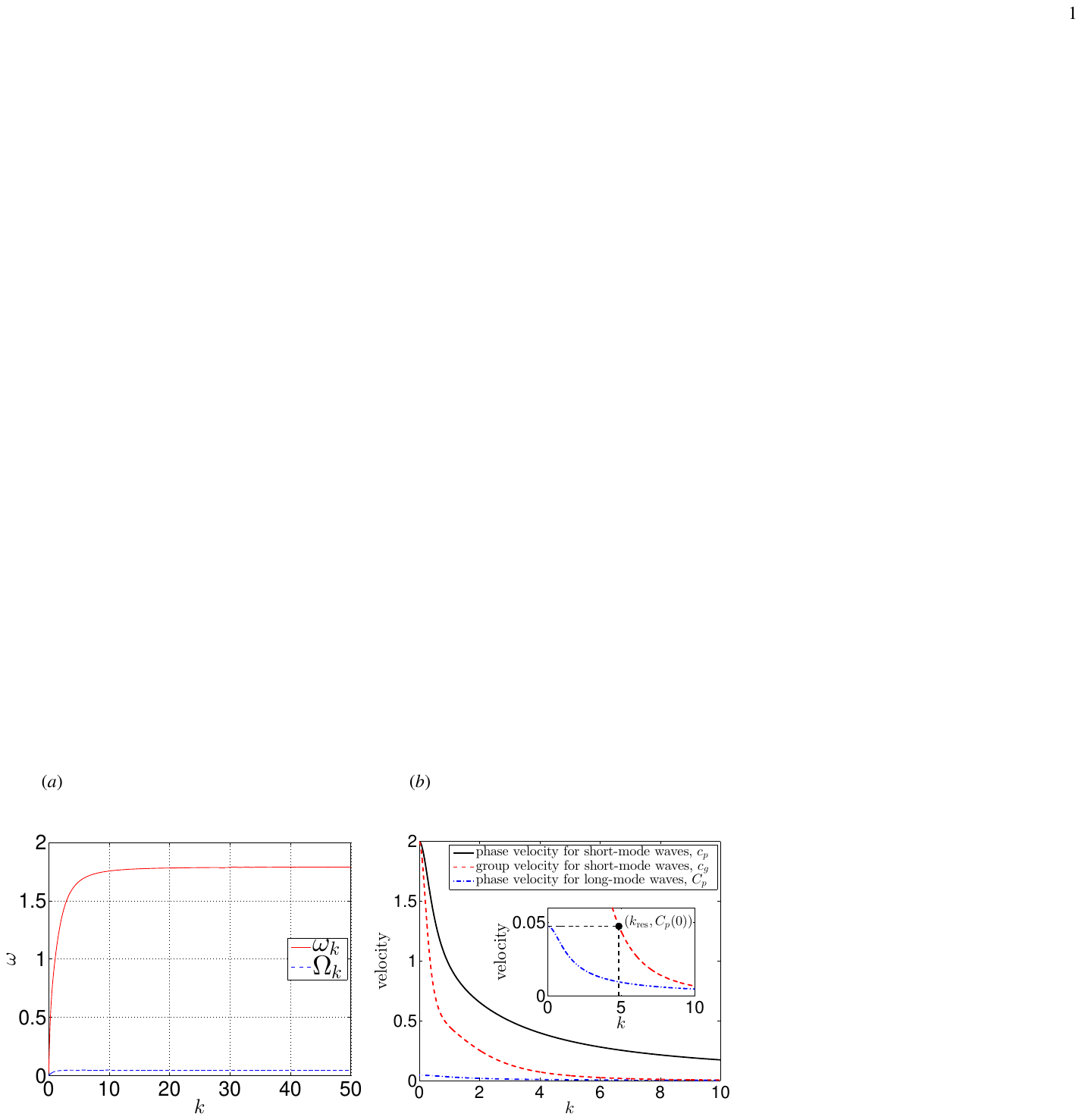}
\caption{(Colour online) (\textit{a}) Pure linear dispersion relations $%
\protect\omega_{k}$ and $\Omega_{k}$ of short-mode waves and long-mode
waves, respectively. (\textit{b}) Comparison of the phase velocity for the
short-mode waves $c_{p}$ (solid line ---------), the group velocity for the
short-mode waves $c_{g}$ (dashed line --\thinspace\ --\thinspace\ --), and
the phase velocity for the long-mode waves $C_{p}$ (dashed-dotted line
--\thinspace $\cdot $\thinspace --\thinspace $\cdot $\thinspace --). The
resonant wavenumber $k_{\mathrm{res}}$ satisfies the triad resonance
condition (\protect\ref{Eqn:resonce_cond}). For our simulations in figures
\protect\ref{Fig:bare_dispersion}--\protect\ref{Fig:OriginSetup_Amplitude},
the parameters are $(h_{1},h_{2},g,\protect\rho _{1},\protect\rho %
_{2})=(1,3,1,1,1.003)$.}
\label{Fig:bare_dispersion}
\end{figure}

The TWN system admits a linear dispersion relation corresponding to a
monochromatic wavetrain with infinitesimal amplitude. By substituting the
monochromatic solutions $(\xi _{j},\overline{u}_{j})\sim \exp [\text{i}
(kx-\mu _{k}t)]$, $j=1,2$, into the TWN system, we obtain the pure linear dispersion relation
between the frequency $\mu _{k}$ and the wavenumber $k$  as

\begin{align}
& \left( 1+\rho _{r}k^{2}h_{1}h_{2}+\frac{1}{3}k^{2}h_{1}^{2}+\frac{1}{3}%
k^{2}h_{2}^{2}+\frac{1}{9}k^{4}h_{1}^{2}h_{2}^{2}+\frac{1}{12}\rho
_{r}k^{4}h_{1}^{3}h_{2}\right) \mu _{k}^{4}  \label{Eqn:dispersion_absence}
\\
& -\left( gh_{1}+gh_{2}+\frac{1}{3}gk^{2}h_{1}h_{2}^{2}+\frac{1}{3}%
gk^{2}h_{1}^{2}h_{2}\right) k^{2}\mu _{k}^{2}+\left( 1-\rho _{r}\right)
g^{2}k^{4}h_{1}h_{2}=0.  \notag
\end{align}%
The same dispersion relation can be found in \citet{Barros2007SAM}. Equation
(\ref{Eqn:dispersion_absence}) always has $4$ real roots when the problem is
considered in the oceanic regime, \textit{i.e.}, when the density ratio $%
\rho _{r}$ is close to $1$. Then, at the leading order in $1-\rho_{r}$, we
can approximate the dispersion relation of the two-mode waves, denoted by $%
\Omega _{k}$ and $\omega _{k}$ [figure \ref{Fig:bare_dispersion}(\textit{a}%
)], as
\begin{equation}
\Omega _{k}^{2}=\frac{\left( 1-\rho _{r}\right) g^{2}h_{1}h_{2}k^{2}}{%
gh_{1}+gh_{2}+\frac{1}{3}gk^{2}h_{1}h_{2}^{2}+\frac{1}{3}gk^{2}h_{1}^{2}h_{2}%
},  \label{Eqn:disp_slow}
\end{equation}%
and
\begin{equation}
\omega _{k}^{2}=\frac{\left( gh_{1}+gh_{2}+\frac{1}{3}gk^{2}h_{1}h_{2}^{2}+%
\frac{1}{3}gk^{2}h_{1}^{2}h_{2}\right) k^{2}}{1+\rho _{r}k^{2}h_{1}h_{2}+%
\frac{1}{3}k^{2}h_{1}^{2}+\frac{1}{3}k^{2}h_{2}^{2}+\frac{1}{9}%
k^{4}h_{1}^{2}h_{2}^{2}+\frac{1}{12}\rho _{r}k^{4}h_{1}^{3}h_{2}}.
\label{Eqn:disp_fast}
\end{equation}%
In the following, the two kinds of waves, corresponding to the dispersion
relations $\Omega _{k}$ [equation (\ref{Eqn:disp_slow})] and $\omega _{k}$ [equation (\ref%
{Eqn:disp_fast})], are referred to as the long-mode (baroclinic) waves and the short-mode
(barotropic) waves, respectively.

Resonant interactions
among different modes are an important process of energy exchange.
For the two-mode waves in equations (\ref{Eqn:disp_slow}) and (\ref{Eqn:disp_fast}), we can numerically verify the existence of
solutions to the class 3 triad resonance condition.  If the
dispersion relations allow the wavenumbers $k_{j}(j=1,2,3)$ and the frequencies
$\omega _{k}(k_{1})$, $\omega _{k}(k_{2})$, and $\Omega _{k}(k_{3})$ to
satisfy the condition
\begin{align}
k_{1}& =k_{2}+k_{3},  \label{triad_bare} \\
\omega _{k}(k_{1})& =\omega _{k}(k_{2})+\Omega _{k}(k_{3}),  \notag
\end{align}%
the corresponding three waves constitute a class 3 resonant triad. Moreover,
if the wavenumbers have the specific form $k_{1}=k_{\text{res}}+\Delta k/2$,
$k_{2}=k_{\text{res}}-\Delta k/2$, $k_{3}=\Delta k$, where $\Delta k\ll k_{%
\text{res}}$ and $\Delta k\rightarrow 0$, then the resonance condition
reduces to
\begin{equation}
c_{g}(k_{\text{res}})=C_{p}(0),  \label{Eqn:resonce_cond}
\end{equation}%
where the group velocity $c_{g}$ and the phase velocity $C_{p}$ are given by
the equations
\begin{equation}
c_{g}(k_{\text{res}})\equiv \left. \frac{\partial \omega _{k}}{\partial k}%
\right\vert _{k=k_{\text{res}}},\ \ C_{p}(0)\equiv \left. \frac{\Omega
_{k}(\Delta k)}{\Delta k}\right\vert _{\Delta k\rightarrow 0}.
\label{Eqn:res_define}
\end{equation}%
Here, $k_{\text{res}}$ is referred to as the \textit{resonant wavenumber}.
Many early results
\citep{Phillips1974ARFM,
Hashizume1980JPSJp,Craig2011NatHazd,Craig2012JFM} have confirmed that there
exists a unique resonant wavenumber $k_{\text{res}}$\ satisfying equation (\ref%
{Eqn:resonce_cond}) in the two-layer system. The resonance condition (\ref%
{Eqn:resonce_cond}) shows that, for resonantly-interacting waves, the group
velocity of short-mode waves and the phase velocity of long-mode waves are
equal
\citep{Phillips1974ARFM,
Hashizume1980JPSJp,Craig2011NatHazd,Craig2012JFM}. Figure \ref%
{Fig:bare_dispersion}(\textit{b}) shows the phase velocity of the short-mode
waves $c_{p}=\omega _{k}/k$, the group velocity of the short-mode waves $%
c_{g}$, and the phase velocity of the long-mode waves $C_{p}=\Omega _{k}/k$.
It can be seen from the inset of figure \ref{Fig:bare_dispersion}(\textit{b}%
) that there indeed exists a resonant wavenumber $k_{\text{res}}$,
satisfying the resonance condition (\ref{Eqn:resonce_cond}), $c_{g}(k_{\text{%
res}})=C_{p}(0)$. Thus, class 3 resonant triads exist among two short-mode
waves and one long-mode wave for the TWN model. Note that the class 3 triad
resonance condition (\ref{triad_bare}) or (\ref{Eqn:resonce_cond}) is not
assumed during the derivation of the TWN model.
Nevertheless, the resonance
condition (\ref{triad_bare}) from \cite%
{Alam2012JFM} can be satisfied using the dispersion relations (\ref{Eqn:disp_slow})
and (\ref{Eqn:disp_fast}) in the TWN model.
As discussed in {\S \S\ \ref%
{sec:nonlinear_class3}, the class 3 triad resonance phenomena will be numerically investigated
in our model in detail.}


In our numerical simulations, throughout the text, the variables are all
dimensionless. That is, numerically, the characteristic length in both
height and wavelength is $h_{1}$, the characteristic speed is $\sqrt{gh_{1}}$%
, and the characteristic time is $\sqrt{h_{1}/g}$ [see similar dimensionless forms used in
numerical simulations in \citep{Choi1999JFM,Choi2009JFM}]. In this work, two parameter regimes
are used in our simulations, one being $%
(h_{1},h_{2},g,\rho _{1},\rho _{2})=(1,3,1,1,1.003)$ and the other being $%
(h_{1},h_{2},g,\rho _{1},\rho _{2})=(1,5,1,0.856,0.996)$.

\subsection{Modified two-layer weakly nonlinear (mTWN) model}

\label{subsec:modif}


Next, we propose a modified version of the TWN model,
which takes into account the effects of a background baroclinic flow.
The basic idea is that we are given a
slowly-varying, sufficiently-long-wavelength background baroclinic flow (BBF), and then study
the behavior of short-mode surface waves on top of this BBF (more details concerning using the BBF
can be found in appendix \ref{subsec:limitation}).
We construct the
modified model for the effects of the BBF on the propagation of short-mode
waves, based on the following assumptions: The long-wave approximation (\ref{Eqn:1.assm_beta}) is only
applied to the short-mode (barotropic) waves of IWs and SWs. The characteristic wavelength of
short-mode waves, $l$, is much shorter than the wavelength of the BBF, $L$,
\begin{equation}
2\pi l/L=\gamma \rightarrow 0.  \label{Eqn:ass_lL}
\end{equation}%
The displacements associated with the BBF at the interface and at the surface
are denoted by $\Xi _{2}^{\ast }$ and $\Xi _{1}^{\ast }$, respectively.
The depth-averaged horizontal currents of the upper and lower layers
are denoted by $\overline{U}_{1}^{\ast }$ and $\overline{U}_{2}^{\ast }$, respectively.
In the rest of the paper, the vector $(\Xi _{j}^{\ast
},\overline{U}_{j}^{\ast })$ is referred to as the BBF.
The slowly-varying long-wavelength BBF disperses weakly, and thereafter the displacements and
currents associated with the BBF, $\Xi _{j}^{\ast }$ and $\overline{U}%
_{j}^{\ast }$, can be expressed as the right-moving
traveling wave $\Xi _{j}^{\ast }(\cdot)=\Xi _{j}^{\ast }\left( K\left( x-C_{s}t\right) \right)$ and %
$\overline{U}_{j}^{\ast }(\cdot)=\overline{U}_{j}^{\ast }\left( K\left( x-C_{s}t\right) \right) $, where $%
K=2\pi /L$ is the characteristic wavenumber and $C_{s}$ is the constant
phase velocity of the BBF. The BBF, $(\Xi _{j}^{\ast
},\overline{U}_{j}^{\ast })$,\ satisfies the sign constraint $\Xi _{1}^{\ast
}>0$, $\Xi _{2}^{\ast }<0$, $\overline{U}_{1}^{\ast }>0$, and $\overline{U}%
_{2}^{\ast }<0$. 
The variables $(\xi _{j},\overline{u}_{j})$ in the TWN model eventually will
be represented as a sum of two components, one being the BBF,
$(\Xi _{j}^{\ast },\overline{U}_{j}^{\ast })$,\ and the
other the perturbations of the waves, $(\xi _{j}^{\ast },\overline{u}%
_{j}^{\ast })$. {We study the motion of the perturbation waves $(\xi
_{j}^{\ast },\overline{u}_{j}^{\ast })$, in particular, we focus on the
dynamical behavior of IWs and SWs on the BBF.}
The assumption (\ref{Eqn:ass_lL}) adopted here is similar to the one adopted
in \citep{Hwung2009JFM}. By substituting $(\xi _{j},\overline{u}_{j})=(\Xi
_{j}^{\ast },\overline{U}_{j}^{\ast })+(\xi _{j}^{\ast },\overline{u}%
_{j}^{\ast })$ into the TWN system, applying asymptotic analysis [see
appendix \ref{subsec:limitation} for details], and casting the system in the
conservation form in the right-moving frame with velocity $C_{s}$, we
obtain the effective evolution equations for the perturbations of waves $%
(\xi _{j}^{\ast },\overline{u}_{j}^{\ast })$\ on the BBF
in dimensional form as follows [referred to as modified two-layer
weakly nonlinear model or mTWN model]: \label{M2_ref}
\begin{equation}
\eta _{1T}^{\ast }-C_{s}\eta _{1X}^{\ast }+(H_{1}^{\ast }\overline{u}%
_{1}^{\ast }+\overline{U}_{1}^{\ast }(\xi _{1}^{\ast }-\xi _{2}^{\ast
})+\vartheta^{\ast } (\xi _{1}^{\ast }-\xi _{2}^{\ast })\overline{u}_{1}^{\ast
})_{X}=0,  \label{Eqn:eff_1}
\end{equation}%
\begin{equation}
\eta _{2T}^{\ast }-C_{s}\eta _{2X}^{\ast }+(H_{2}^{\ast }\overline{u}%
_{2}^{\ast }+\overline{U}_{2}^{\ast }\xi _{2}^{\ast }+\vartheta^{\ast } \xi
_{2}^{\ast }\overline{u}_{2}^{\ast })_{X}=0,  \label{Eqn:eff_2}
\end{equation}%
\begin{equation}
M_{1T}^{\ast }-C_{s}M_{1X}^{\ast }+\left( \overline{U}_{1}^{\ast }\overline{u%
}_{1}^{\ast }+g\xi _{1}^{\ast }+\frac{1}{2}\vartheta^{\ast } \overline{u}_{1}^{\ast
2}\right) _{X}+P_{1X}/\rho _{1}=0,  \label{Eqn:eff_3}
\end{equation}%
\begin{equation}
M_{2T}^{\ast }-C_{s}M_{2X}^{\ast }+\left( \overline{U}_{2}^{\ast }\overline{u%
}_{2}^{\ast }+g\xi _{2}^{\ast }+\rho _{r}g\eta _{1}^{\ast }+\frac{1}{2}%
\vartheta^{\ast } \overline{u}_{2}^{\ast 2}\right) _{X}+P_{1X}/\rho _{2}=0.
\label{Eqn:eff_4}
\end{equation}%
Here%
\begin{equation*}
H_{1}^{\ast }=h_{1}+\Xi _{1}^{\ast }-\Xi _{2}^{\ast },\text{ \ \ }%
H_{2}^{\ast }=h_{2}+\Xi _{2}^{\ast },
\end{equation*}%
\begin{equation*}
\eta _{1}^{\ast }=h_{1}+\xi _{1}^{\ast }-\xi _{2}^{\ast },\text{ \ \ }\eta
_{2}^{\ast }=h_{2}+\xi _{2}^{\ast },
\end{equation*}%
\begin{equation*}
M_{1}^{\ast }=\overline{u}_{1}^{\ast }-\frac{1}{3}h_{1}^{2}\overline{u}%
_{1XX}^{\ast }-\frac{1}{2}h_{1}\left( H_{2}^{\ast }\overline{u}_{2}^{\ast }+%
\overline{U}_{2}^{\ast }\xi _{2}^{\ast }+\vartheta^{\ast } \overline{u}_{2}^{\ast
}\xi _{2}^{\ast }\right) _{XX},
\end{equation*}%
\begin{equation*}
M_{2}^{\ast }=\overline{u}_{2}^{\ast }-\frac{1}{2}\rho _{r}h_{1}^{2}%
\overline{u}_{1XX}^{\ast }-\frac{1}{3}h_{2}^{2}\overline{u}_{2XX}^{\ast
}-\rho _{r}h_{1}\left( H_{2}^{\ast }\overline{u}_{2}^{\ast }+\overline{U}%
_{2}^{\ast }\xi _{2}^{\ast }+\vartheta^{\ast } \overline{u}_{2}^{\ast }\xi
_{2}^{\ast }\right) _{XX},
\end{equation*}%
and the right-moving frame corresponds to the variable transformation $T=t$
and $X=x-C_{s}t$, where $C_{s}$ is the phase velocity of the BBF
$(\Xi _{j}^{\ast },\overline{U}_{j}^{\ast })$, and $P_{1}$ is the
normal pressure exerted at the surface. The parameter $\vartheta^{\ast } $ takes two
values 1 or 0, where $\vartheta^{\ast } =0$ corresponds to the linear regime and $%
\vartheta^{\ast } =1$ corresponds to the nonlinear regime for the perturbation waves
$(\xi _{j}^{\ast },\overline{u}_{j}^{\ast })$. In the right-moving frame,
the BBF $(\Xi _{j}^{\ast },\overline{U}_{j}^{\ast })$
is steady with respect to time $T$. More details concerning the dimensionless
form of the mTWN model can be found in appendix \ref{subsec:limitation}.


\section{Review of linear modulation theory and nonlinear class 3 triad
resonance}

\label{sec:review_lin_non}

\subsection{Linear modulation (or ray-based) theory}

\label{sec:linear_modul}

In this section, we review the linear modulation (or ray-based) theory for
short-mode SWs on a BBF. We take the components of
the long-wavelength BBF, $(\Xi _{1}^{\ast },\Xi _{2}^{\ast },\overline{U}%
_{1}^{\ast },\overline{U}_{2}^{\ast })$, to be proportional to
\begin{equation}
\left\{ \tanh \left[ s_{\Xi ^{\ast }}\left( X+X_{\Xi ^{\ast }}\right) %
\right] -\tanh \left[ s_{\Xi ^{\ast }}\left( X-X_{\Xi ^{\ast }}\right) %
\right] \right\} /2,  \label{baroclinic_flow_form}
\end{equation}%
with the maximal amplitudes of $(\Xi _{1}^{\ast },\Xi _{2}^{\ast },\overline{U}%
_{1}^{\ast },\overline{U}_{2}^{\ast })$ being $(0.0003,-0.1,0.007,-0.006)$,
the slope $s_{\Xi ^{\ast }}=0.2$, the wavelength $X_{\Xi ^{\ast }}=120$, and
the phase velocity $C_{s}=0.0485$. Here, the maximal amplitudes and the phase velocity are
chosen to be close to those of the interfacial solitary wave solution of the TWN model for $\Xi_{2}
^{\ast }(0)=-0.1$ [see appendix \ref{subsec:traveling_wave}]. The parameter $%
X_{\Xi ^{\ast }}=120$ characterizes the wavelength of the BBF and is
chosen to be large enough to satisfy the assumption (\ref%
{Eqn:ass_lL}). The parameter $s_{\Xi ^{\ast }}$ characterizes the slope of
the BBF. Later on in \S \S\ \ref{sec:NESS_setting}, we will vary these
parameters for a parametric study and see how their values affect the
dynamical behavior of short-mode SWs.

In the presence of the baroclinic flow
in the right-moving frame, the modulated dispersion relation $\overline{\nu }%
_{k}$ for the perturbation waves $(\xi _{j}^{\ast },\overline{u}_{j}^{\ast })$
can be obtained by substituting $(\xi _{j},\overline{u}_{j})\sim (\Xi
_{j}^{\ast },\overline{U}_{j}^{\ast })+\exp [\text{i}(kX-\overline{\omega }%
_{k}T)]$ with $\overline{\omega }_{k}=\overline{\nu }_{k}+C_{s}k$ into the
TWN system {}. Here, $\overline{\omega }_{k}$ is the dispersion relation in
the resting frame, whereas $\overline{\nu }_{k}$\ is the dispersion relation
in the right-moving frame. The resulting equation is
\begin{equation}
\left\vert
\begin{array}{cccc}
\begin{smallmatrix}
-\overline{\omega }_{k}+k\overline{U}_{1}^{\ast } & \overline{\omega }_{k}-k%
\overline{U}_{1}^{\ast } & kh_{1}+k\Xi _{1}^{\ast }-k\Xi _{2}^{\ast } & 0 \\
0 & -\overline{\omega }_{k}+k\overline{U}_{2}^{\ast } & 0 & kh_{2}+k\Xi
_{2}^{\ast } \\
gk & -\frac{1}{2}kh_{1}\overline{\omega }_{k}^{2} & -\overline{\omega }_{k}-%
\frac{1}{3}k^{2}h_{1}^{2}\overline{\omega }_{k}+k\overline{U}_{1}^{\ast } & 0
\\
\rho _{r}gk & (1-\rho _{r})gk-\rho _{r}kh_{1}\overline{\omega }_{k}^{2} & -%
\frac{1}{2}\rho _{r}k^{2}h_{1}^{2}\overline{\omega }_{k} & -\overline{\omega
}_{k}-\frac{1}{3}k^{2}h_{2}^{2}\overline{\omega }_{k}+k\overline{U}%
_{2}^{\ast }%
\end{smallmatrix}
&  &  &
\end{array}%
\right\vert =0,  \label{modulated_DR}
\end{equation}%
where $\left\vert \cdot \right\vert $ denotes the determinant of the
enclosed matrix. Moreover, we also obtain three linear eigenfunction
relations among the amplitudes of $(\xi _{j}^{\ast },\overline{u}_{j}^{\ast
})$. We note that, first, $(\Xi _{j}^{\ast },\overline{U}_{j}^{\ast })$ are
steady in time $T$ in the right-moving frame, so that the dispersion
relation $\overline{\nu }_{k}$\ is independent of time $T$. Second, the
wavelengths of $(\Xi _{j}^{\ast },\overline{U}_{j}^{\ast })$ are relatively
long compared to the characteristic length of the short-mode waves, so $(\Xi _{j}^{\ast },\overline{U}%
_{j}^{\ast })$ in the dispersion relation (\ref{modulated_DR}) can be
locally treated as constant in $X$ space. Third, if $(\Xi _{j}^{\ast },%
\overline{U}_{j}^{\ast })=0$, the modulated dispersion relation $\overline{%
\omega }_{k}$ in equation (\ref{modulated_DR}) reduces to the pure linear dispersion
relation $\mu _{k}$ in equation (\ref{Eqn:dispersion_absence}).


Due to the slow variation in space and time of the phase of the short-mode
waves, the governing equations of the space-time rays at the location $X$
and wavenumber $k$ are given by the ray-based theory %
\citep{Whitham1974Linear,Bakhanov2002JGR,Kodaira2016JFM},%
\begin{equation}
\frac{dX}{dT}=\frac{\partial \overline{\nu }_{k}}{\partial k},
\label{Eqn:ray_base}
\end{equation}%
\begin{equation}
\frac{dk}{dT}=-\frac{\partial \overline{\nu }_{k}}{\partial X},
\label{Eqn:wavenum_conserve}
\end{equation}%
where $\overline{\nu }_{k}$ is the modulated dispersion relation in equation (\ref%
{modulated_DR}). Since the dispersion relation $\overline{\nu }_{k}$ does
not explicitly depend on time $T$, equations (\ref{Eqn:ray_base}) and (\ref%
{Eqn:wavenum_conserve}) constitute a Hamiltonian system with $\overline{\nu }%
_{k}$ as the Hamiltonian, $X$ displacement, and $k$ momentum. Equation (\ref%
{Eqn:ray_base}) states that the wave packet propagates with the group velocity.

\begin{figure}
\center 
\includegraphics[scale=1.05]{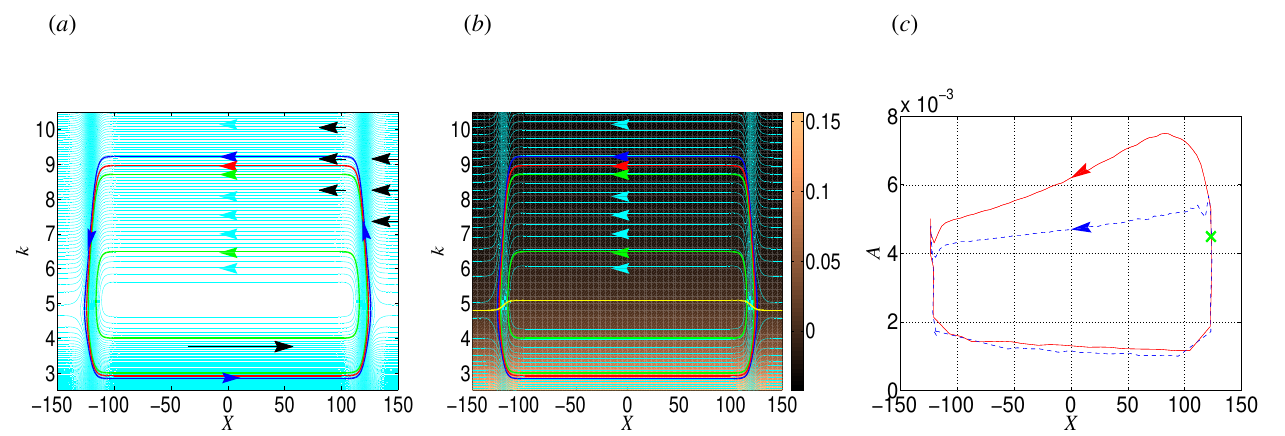}
\caption{(Colour online) (\textit{a}) The phase portrait of the right-moving wave-packet
motion in the variables $X$ and $k$. Along each cyan-colored curve, the
frequency $\overline{\protect\nu }_{k}$ remains constant. Inside the region
enclosed by the blue level-curve, the right-moving short-mode waves are trapped. The red
level-curve corresponds to the simulation in panel (\textit{c}). Along the
two green level-curves, the minimal wavenumbers are $k = 3$ and $k = 4$,
respectively [corresponding to the initial perturbation (\protect\ref%
{initial_force}) in the next \S\ \ref{sec:strong_nonlinear}]. The arrows indicate the direction of wave-packet
movement. (\textit{b}) Plotted is the group velocity $\partial \overline{%
\protect\nu }_{k}/{\partial k}$ with its magnitude color-coded
[copper-black] on the background. The yellow curve corresponds to zero group
velocity. (\textit{c}) The maximal amplitude $A$ versus the peak location $X$
for the evolution of a narrow wave packet with a narrow band of wavenumbers.
The initial value is marked by the green cross. The red curve corresponds to
the first period and the blue curve corresponds to the second period. {The
main observation is that SWs become high at the leading edge, $100<X<120$,
and low at the trailing edge, $-120<X<-100$ (asymmetric behavior \hyperref[asym3%
]{[1c]}). This main observation of asymmetric behavior \hyperref[asym3]{[1c]}
persists for both periods (red and blue curves).
On the flat BBF ($-100<X<100$), as a result of the wave dispersion, the
initially narrow SW packet becomes smaller in maximal amplitude and wider
in length. Also, due to the wave
dispersion, the maximal amplitude in the second period (blue curve) becomes
smaller than that in the first period (red curve). In fact, eventually the
waves will be almost uniformly distributed on the flat BBF after several periods
[not shown here] due to the wave dispersion and the periodic motion of SWs.} }
\label{Fig:Broad_near_field_theory}
\end{figure}


We first study the dynamical behavior of right-moving (positive phase velocity) short-mode SWs.
In figure \ref{Fig:Broad_near_field_theory}(\textit{a}), the phase portrait
of the right-moving wave-packet motion in the variables $X$ and $k$ is constructed from the
modulated dispersion relation $\overline{\nu }_{k}$ using the ray-based
theory. It can be seen from figure \ref{Fig:Broad_near_field_theory}(\textit{%
a}) that right-moving short-mode waves become short in wavelength at the leading edge ($%
X>0$), whereas they become long at the trailing edge ($X<0$). Figure \ref%
{Fig:Broad_near_field_theory}(\textit{b}) displays the group velocity of these
short-mode waves, $\partial \overline{\nu }_{k}/\partial k$, (in variables $%
X $ and $k$) with its magnitude color-coded on top of the background. It can
be seen from figure \ref{Fig:Broad_near_field_theory}(\textit{b}) that,
above the yellow curve, $\partial \overline{\nu }_{k}/\partial k<0$ and its
absolute value is relatively small, whereas below the yellow curve, $%
\partial \overline{\nu }_{k}/\partial k>0$ and its absolute value is
relatively large. This suggests that the right-moving short-mode wave packets propagate
towards the trailing edge [in the direction of decreasing $X$ when $C_{s}>0$%
] with a relatively small group velocity, and towards the leading edge [in
the direction of increasing $X $ when $C_{s}>0$] with a relatively large
group velocity.

We now study the relation between the maximal amplitude $A$ and the peak
location $X$ of the right-moving short-mode waves. We numerically simulate the mTWN model
{} with $\vartheta^{\ast} =1$ and $P_{1}=0$ 
in the weakly nonlinear regime. Here, the weakly nonlinear regime means that
the amplitude of initial perturbation $(\xi _{j}^{\ast },\overline{u}%
_{j}^{\ast })$ is very small. The initial perturbation for short-mode waves
is a narrow wave packet with a disturbance of a narrow band of wavenumbers.
The initial perturbation of $\xi _{2}^{\ast }$ is taken to be
\begin{equation}
\xi _{2}^{\ast }=A_{\varepsilon }[\tanh (X-X_{0}+x_{0})-\tanh
(X-X_{0}-x_{0})]\cos (k_{0}(X-X_{0})),  \label{Eqn:perturb}
\end{equation}%
with $A_{\varepsilon }=2\times 10^{-4}$, $X_{0}=123$, $x_{0}=3$ and $k_{0}=5$%
. This initial perturbation $(X_{0},k_{0})$ is located on the red-level
curve in figure \ref{Fig:Broad_near_field_theory}(\textit{a}). To allow for
only right-moving short-mode waves in the system, the other three variables,
$(\xi _{1}^{\ast },\overline{u}_{1}^{\ast },\overline{u}_{2}^{\ast })$, have
the same form as the perturbation (\ref{Eqn:perturb}), with their amplitudes
satisfying the linear eigenfunction relations following from the matrix in equation (%
\ref{modulated_DR}). For example, we can obtain from the matrix in equation (\ref%
{modulated_DR}) that $\overline{u}_{2}^{\ast }=\left( \overline{\omega }%
_{k}-k\overline{U}_{2}^{\ast }\right) \xi _{2}^{\ast }/\left( kh_{2}+k\Xi
_{2}^{\ast }\right) $. More details about the settings of the numerical
simulations, including the numerical scheme, dealiasing strategy,
boundary conditions, and Kelvin-Helmholtz instability are presented in
appendix \ref{accuracy_test}.

Figure \ref{Fig:Broad_near_field_theory}(\textit{c}) displays the maximal
amplitude $A$ versus the peak location $X$ for the evolution of the
perturbation for the right-moving short-mode waves. It can be seen that SWs become high
in amplitude at the leading edge ($X>0$) and low at the trailing edge ($X<0$%
). {Notice that this assertion is also true for the Euler system of equations \citep{Lewis1974JFM}.%
}
In particular, this SWs' amplitude phenomenon can be inferred from the energy conservation
equation of the Euler system.
The energy conservation equation can be obtained using the concept of
radiation stress for the Euler system \citep{Lewis1974JFM}:%
\begin{equation}
\frac{\partial \widetilde{A}^{2}}{\partial t}+\frac{\partial \left[ \left(
\widetilde{C}_{g}+\widetilde{U}\right) \widetilde{A}^{2}\right] }{\partial x}%
=-\frac{1}{2}\widetilde{A}^{2}\frac{\partial \widetilde{U}}{\partial x},
\label{eqn:radiation_stress}
\end{equation}%
where $\widetilde{C}_{g}$ is the group velocity in the pure linear
dispersion relation, $\widetilde{A}$ is the amplitude of SW packets, and $%
\widetilde{U}$ is the surface-current elevation. At the leading edge, a
negative strain rate $\partial \widetilde{U}/\partial x$\ corresponds to the
growth of the amplitude [surface convergence], whereas at the trailing edge,
a positive strain rate $\partial \widetilde{U}/\partial x$ corresponds to
the decay of the amplitude [surface divergence].

\begin{figure}
\center 
\includegraphics[scale=0.9]{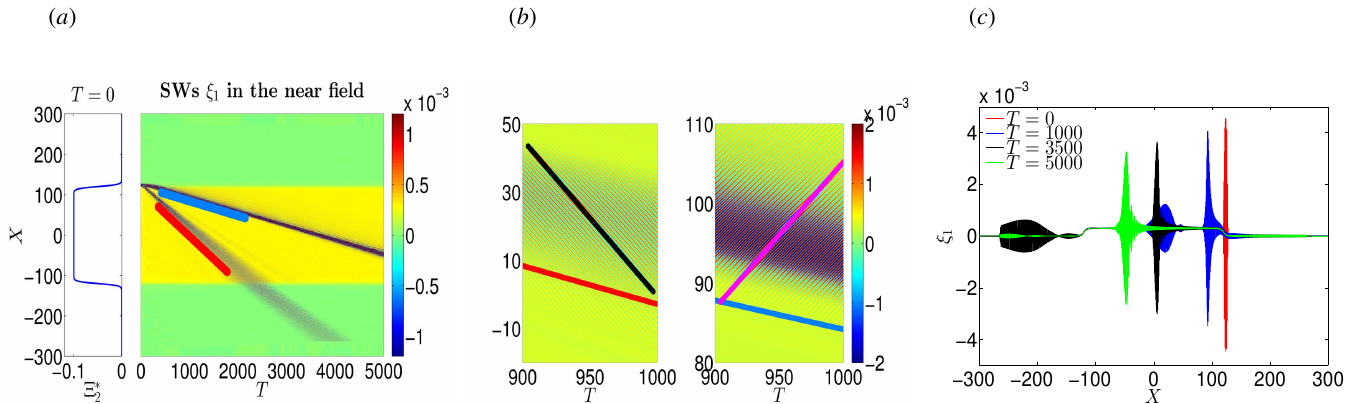}
\caption{(Colour online) (\textit{a}) The snapshot of the BBF (left panel) and
the spatiotemporal evolution of
SWs' profile $\protect\xi _{1}$. The red and blue lines correspond to the group velocities of the left-moving
SWs and right-moving SWs, respectively.
(\textit{b}) Zoomed-in version of (\textit{a}). The red and blue lines are the same as those in (\textit{a}).
The black and magenta lines correspond to the phase velocities of the left-moving SWs and right-moving SWs, respectively.
Note that  the phase velocity of the left-moving SWs [black line] is negative whereas the phase velocity of the right-moving SWs [magenta line] is positive.
(\textit{c}) Snapshots of SWs' profile $\protect\xi _{1}$. Initially at $T=0$ (red curve), there is only a narrow wave packet
of SWs.  At $T=1000$, the smaller-amplitude left-moving SW packet (blue curve for $X<50$) and the larger-amplitude right-moving SW packet (blue curve for $X>50$)
can be seen. Note that the left-moving SW packet, corresponding to a negative phase velocity, is not trapped in the near field.  On the other hand, the right-moving SW packet, corresponding to a positive phase velocity, is trapped in the near field.   At $T=3500$, the left-moving waves (black curve for $X<-150$) are partially absorbed by the boundary condition [see appendix \ref{accuracy_test}]. At $T=5000$, the left-moving waves (green curve for $X<-150$) are completely absorbed by the boundary.
For the right-moving waves, as a result of the wave dispersion, the
initially narrow SW packet (red curve for $X>100$) becomes smaller in maximal amplitude and wider
in length (\textit{e.g.}, green curve for $X>-50$).
}
\label{Fig:LeftGoing}
\end{figure}



Next, we study the dynamical behavior of left-moving (negative phase velocity) short-mode SWs.
Figure \ref{Fig:LeftGoing}(\textit{a}) displays the spatiotemporal evolution of SWs' profile
$\xi_{1}$. The initial perturbation of $\xi_{1}^{\ast }$ is the same as that of $\xi_{1}^{\ast }$ used
in figure \ref{Fig:Broad_near_field_theory}(\textit{c}). However, the initial perturbations of
the other three variables, $\xi_{2}^{\ast},u_{1}^{\ast},u_{2}^{\ast}$, are all taken to be zero.
It can be seen that the initial narrow wave packet splits into two wave packets,
one corresponding to the right-moving SW packet [blue line] and the other corresponding to the
left-moving SW packet [red line]. The right-moving waves are trapped
in the near field as discussed above and shown in figure \ref{Fig:Broad_near_field_theory}.
On the other hand, the left-moving waves  quickly leave the near field
[figure \ref{Fig:LeftGoing}],
and then these left-moving waves are absorbed by the boundary
[figures \ref{Fig:LeftGoing}(\textit{a}) and (\textit{c})].


To summarize, the behavior of the short-mode waves is asymmetric at the
leading edge vs. the trailing edge when there is an underlying BBF:

\noindent [1a] \label{asym1}\textit{Right-moving SW packets propagate towards the
trailing edge with a relatively small group velocity, and towards the
leading edge with a relatively large group velocity [figure \ref%
{Fig:Broad_near_field_theory}(b)].}

\noindent [1b] \label{asym2}\textit{Right-moving SWs become short in wavelength at the
leading edge and long at the trailing edge [figure \ref%
{Fig:Broad_near_field_theory}(a)]. }

\noindent [1c] \label{asym3}\textit{Right-moving SWs become high in amplitude at the
leading edge and low at the trailing edge [figure \ref%
{Fig:Broad_near_field_theory}(c)]. }

\noindent [1d] \label{asym4}\textit{Left-moving SWs quickly leave the near field, and then only the right-moving
SWs that propagate in the same direction as the BBF remain trapped in the near field [figure \ref{Fig:LeftGoing}].}

{\ \noindent To the best of our knowledge, the asymmetric behavior \hyperref[asym2%
]{[1b]} and \hyperref[asym4%
]{[1d]} was earlier discovered in references %
\citep{Bakhanov2002JGR,Kodaira2016JFM}, the asymmetric behavior \hyperref[asym3%
]{[1c]} was earlier discovered in reference \citep{Lewis1974JFM}, but the
asymmetric behavior \hyperref[asym1]{[1a]} was not pointed out
earlier. Here, we review these asymmetric types of behavior predicted by the
ray-based theory using the mTWN model 
in the weakly nonlinear regime. These asymmetric types of behavior are important for
demonstrating the modulation-resonance mechanism, as will be discussed in \S\ %
\ref{sec:strong_nonlinear} below.}

In the rest of the paper, we always use short-mode SWs to denote right-moving short-mode SWs
that are trapped in the near field. Whenever we discuss left-moving short-mode SWs, we will directly use the
phrase left-moving short-mode SWs.

\subsection{Nonlinear class 3 triad resonance}

\label{sec:nonlinear_class3}

\begin{figure}
\centering \hspace*{-5mm}
\includegraphics[scale=1.30]{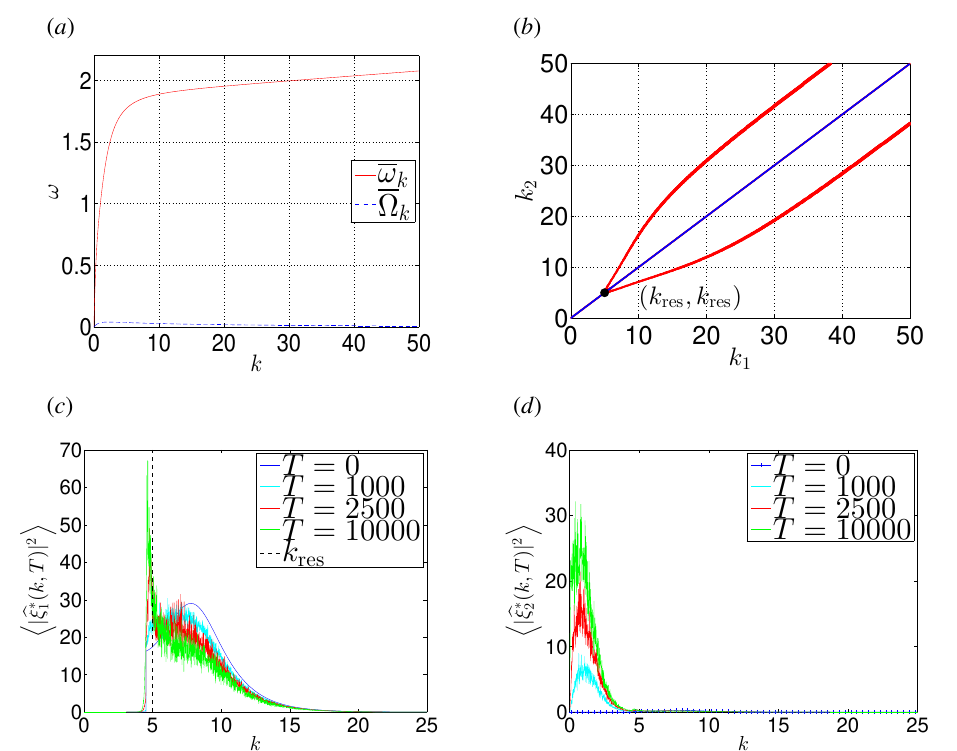}
\caption{(Colour online) (\textit{a}) Modulated dispersion relations (%
\protect\ref{modulated_DR}) for the short-mode waves, $\overline{\protect%
\omega}_{k}$, and the long-mode waves, $\overline{\Omega}_{k}$. (\textit{b})
The wavenumbers of two short-mode waves satisfying the class 3 triad
resonance condition as in equation (\protect\ref{triad_bare}) but using the modulated
dispersion relations. Panels (\textit{c}) and (\textit{d}) show the
evolution of the wavenumber spectra of the SWs $\protect\xi_{1}^{*}$ and the
IWs $\protect\xi_{2}^{*}$. The resonant wavenumber is $k_{\mathrm{res}}=5.0$%
. }
\label{Fig:class_resonance}
\end{figure}

In this section, we review the phenomena associated with the nonlinear class
3 triad resonance for short-mode SWs.
Class 3 triad resonance is considered responsible for the surface
signature of the underlying internal waves. The resonant excitation of surface
waves whose group velocity is close to the phase velocity of long internal
waves has been reported in various observations %
\citep{Lewis1974JFM,Osborne1980Sci}. Many works have been dedicated to the
effective model describing class 3 triad-resonance coupling phenomena
between interfacial and surface waves using multi-scale analysis %
\citep{Lewis1974JFM,Kawahara1975JPSJ,Hashizume1980JPSJp,Funakoshi1983JPSJp,Sepulveda1987PhysFld,Lee2007JKrPS,Hwung2009JFM,Craig2012JFM}%
. \cite{Alam2012JFM} showed that class 3 triad resonance can occur among two
short-mode waves and one long-mode wave in which all waves propagate in the
same direction in a two-layer system. \cite{Tanaka2015JFM} showed that in a
two-layer Euler system, the spectra of SWs and IWs can change significantly
due to the class 3 triad resonance. Moreover, there is an inverse energy
cascade from larger wavenumbers to smaller wavenumbers, and finally a steep
peak forms in the spectrum of SWs around the resonant wavenumber $k_{\text{%
res}}$ \citep{Tanaka2015JFM}.

In our setup, we will reproduce the above phenomena observed by \cite%
{Tanaka2015JFM} using our mTWN model with $\vartheta^{\ast} =1$ and $P_{1}=0$ in
the presence of the BBF 
in the strongly nonlinear regime. Here, the strongly nonlinear regime means
that the amplitude of the initial perturbation of $(\xi _{j}^{\ast },%
\overline{u}_{j}^{\ast })$ is relatively large. The BBF
is constant in $X$ space, $(\Xi _{1}^{\ast },\Xi _{2}^{\ast },\overline{%
U}_{1}^{\ast },\overline{U}_{2}^{\ast })=(0.0003,-0.1,0.007,-0.006)$, with
the velocity $C_{s}=0.0485$. Periodic boundary conditions are applied in
this numerical implementation [note that, for other numerical implementations
in \S \S\ \ref{sec:linear_modul} and \S\ \ref{sec:strong_nonlinear},
absorbing boundary conditions are applied as discussed in appendix \ref%
{accuracy_test}]. The spectrum of the IWs' amplitude $\xi _{2}^{\ast }$, at the initial time
$T=0$, is given by
\begin{equation}
\left\vert \widehat{\xi }_{2}^{\ast }\left( k,0\right) \right\vert ^{2}=a_{m}%
\mathrm{sech}\left[ c_{w}(k-k_{c})\right] H_{e}\left( k-k_{m}\right) ,
\label{initial_IW_absence}
\end{equation}%
where $a_{m}=0.25$, $c_{w}=0.5$, $k_{c}=8$, $k_{m}=4.5$, and $H_{e}\left(
k-k_{m}\right) $ is a Heaviside step function with $H_{e}=0$ for $k\leq
k_{m} $ and $H_{e}=1$ for $k>k_{m}$. Here, the wavenumber $k_{c}>k_{\text{res%
}}$, and the Heaviside step function $H_{e}$ is used to ensure that low
wavenumbers [$k\leq k_{m}$] are not included in the initial disturbance of
the IW spectrum, and furthermore to identify the subsequent growth of the IW
spectrum for $k\leq k_{m}$ due to class 3 triad resonance. To ensure that
the initial wave field consists only of the short-mode waves propagating in
the positive $X$ direction, again, the amplitudes of the other three
variables $(\xi _{1}^{\ast },\overline{u}_{1}^{\ast },\overline{u}_{2}^{\ast
})$ at $T=0$ satisfy the linear eigenfunction relations stemming from the
matrix in equation (\ref{modulated_DR}). The phase of the variable $\widehat{\xi} _{2}^{\ast }\left( k,0\right)$ for each
wavenumber is uniformly distributed on $[0,2\pi ]$.


Figure \ref{Fig:class_resonance}(\textit{a}) displays the modulated
dispersion relation $\overline{\omega }_{k}$ in equation (\ref{modulated_DR}) for the
short-mode waves, and $\overline{\Omega }_{k}$ in equation (\ref{modulated_DR}) for
the long-mode waves, in the presence of the constant BBF.
Figure \ref{Fig:class_resonance}(\textit{b}) displays the wavenumbers $%
k_{1}$ and $k_{2}$ of the two short-mode waves which satisfy the class 3
triad resonance condition as in equation (\ref{triad_bare}), but using the modulated
dispersion relations $\overline{\omega }_{k}$\ in equation (\ref{modulated_DR}) for
two short-mode waves, and $\overline{\Omega }_{k}$ in equation (\ref{modulated_DR})
for one long-mode wave. Figures \ref{Fig:class_resonance}(\textit{c}) and (%
\textit{d}) display several snapshots of the wave spectra for SWs and IWs,
respectively.

It can be clearly seen that

\noindent [2a] \label{creso1} \textit{a significant amount of energy is
transferred from SWs towards IWs due to the class 3 triad resonance}.

\noindent [2b] \label{creso2} \textit{There is an inverse energy cascade for
the spectrum of SWs from larger wavenumbers above }$k_{\text{res}}$\textit{\
towards smaller wavenumbers, and finally a steep peak forms in the spectrum
of SWs around }$k_{\text{res}} $.

{Here, we have reproduced the class 3 triad resonance phenomena (as
observed by \cite{Tanaka2015JFM}) using our mTWN model. These two
resonance phenomena \hyperref[creso1]{[2a]} and \hyperref[creso2]{[2b]} are also important for demonstrating the
modulation-resonance mechanism 
in \S\ \ref{sec:strong_nonlinear} below. }



\section{Modulation-resonance mechanism for resonant surface waves at the
leading edge}

\label{sec:strong_nonlinear}

We now address the main question of how SWs behave in $X$ and $k$ spaces
in the presence of an underlying slowly-varying long-wavelength BBF.
We will see that our numerical results exhibit surface signatures
qualitatively consistent with the observed surface phenomena as mentioned in
introduction:
\hyperlink{ASYM_MODU}{(i)}
the surface ripples at the leading
edge, and
\hyperlink{CLASS_RESON}{(ii)}
resonant excitations of these
surface ripples. We then propose a possible mechanism underlying these
surface phenomena.

\subsection{Modulation-resonance mechanism}

\label{sec:modu_res}

\begin{figure}
\centering \hspace*{-10mm} %
\includegraphics[width=5.0in]{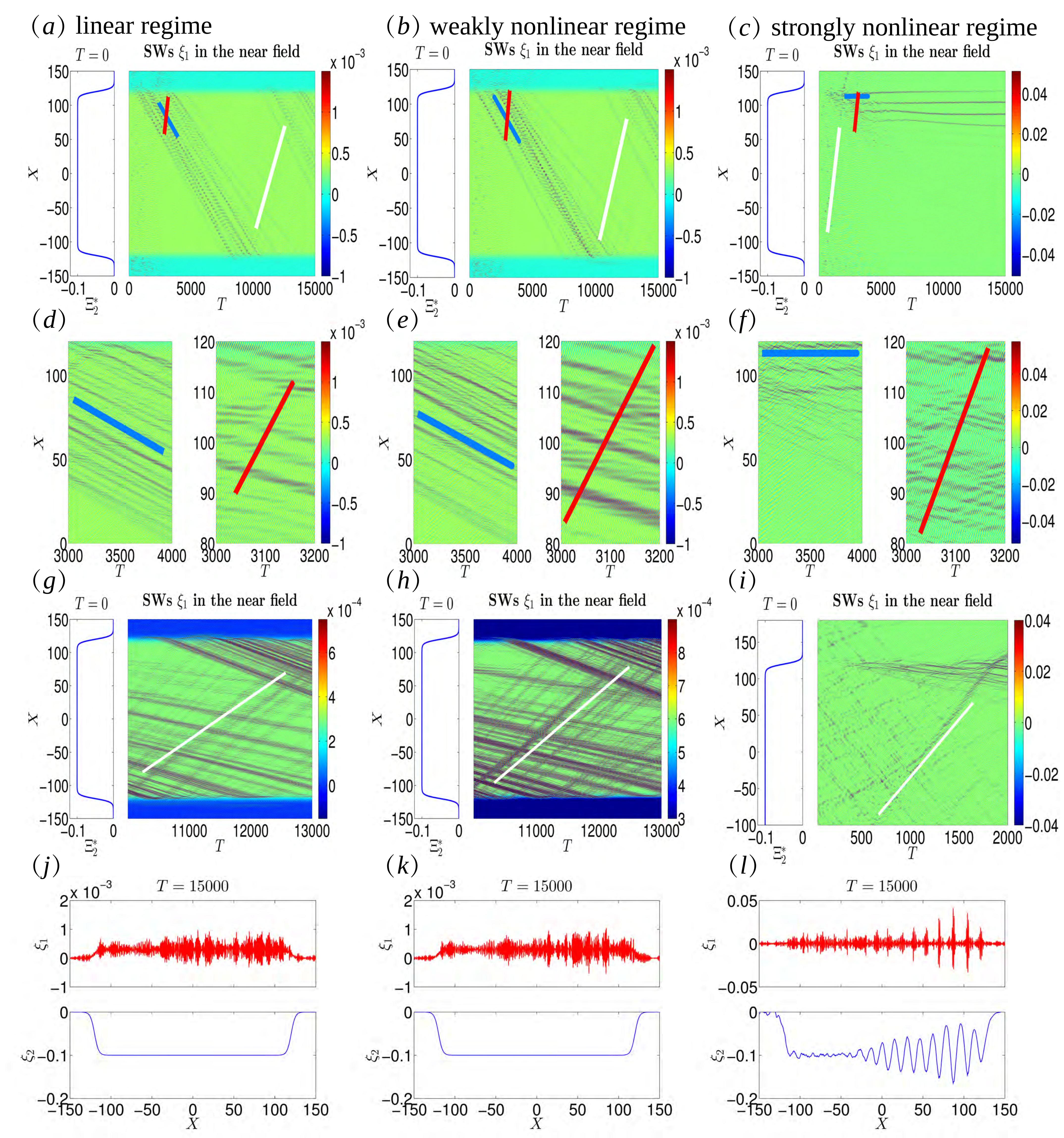}
\caption{(Colour online) Comparison of spatiotemporal manifestation of SWs in the
linear regime ($\protect\vartheta^{\ast}=0,\protect\sigma_{m}=5$E$-4$) [first
column], in the weakly nonlinear regime ($\protect\vartheta^{\ast}=1,\protect\sigma%
_{m}=5$E$-4$) [second column], and in the strongly nonlinear regime ($%
\protect\vartheta^{\ast}=1,\protect\sigma_{m}=2.5$E$-2$) [third column]. The first
row [(\textit{a})-(\textit{c})] displays  the snapshot of the BBF (left panel) and  the spatiotemporal evolution of
SWs' profile $\protect\xi _{1}=\Xi _{1}^{*}+\protect\xi _{1}^{*}$ (right panel).
The parameters
of the BBFs are given in table \ref{tab:regime1}.
The second row [(\textit{d})-(\textit{f})] and the third row [(\textit{g})-(\textit{i})] display the zoomed-in versions of
SWs' profiles. The red line corresponds to
the positive phase velocity of the right-moving short-mode SWs.
The white line (blue line) corresponds to the group velocity of right-moving short-mode SWs when SW packets
propagate from the trailing (leading) edge towards the leading (trailing) edge. The ranges of color bars in panels (\textit{a})-(\textit{f})
are  automatically tuned. To observe the small-amplitude SW packets that propagate
from the trailing edge towards the leading edge [white lines], the ranges of color bars in
panels (\textit{g})-(\textit{i}) are manually tuned to be smaller.
The
fourth row [(\textit{j})-(\textit{l})] displays the snapshots of SWs and IWs
at $T = 15000$.
It can be seen that spatiotemporal manifestation of SWs differs greatly between
the linear (or weakly nonlinear) regime and the strongly nonlinear regime. In
the strongly nonlinear regime, SW packets are located at the leading edge due to the class 3 triad resonance [see text]. }
\label{Fig:Background_Strong}
\end{figure}

\begin{figure}
\centering \hspace*{-10mm} %
\includegraphics[width=5.0in]{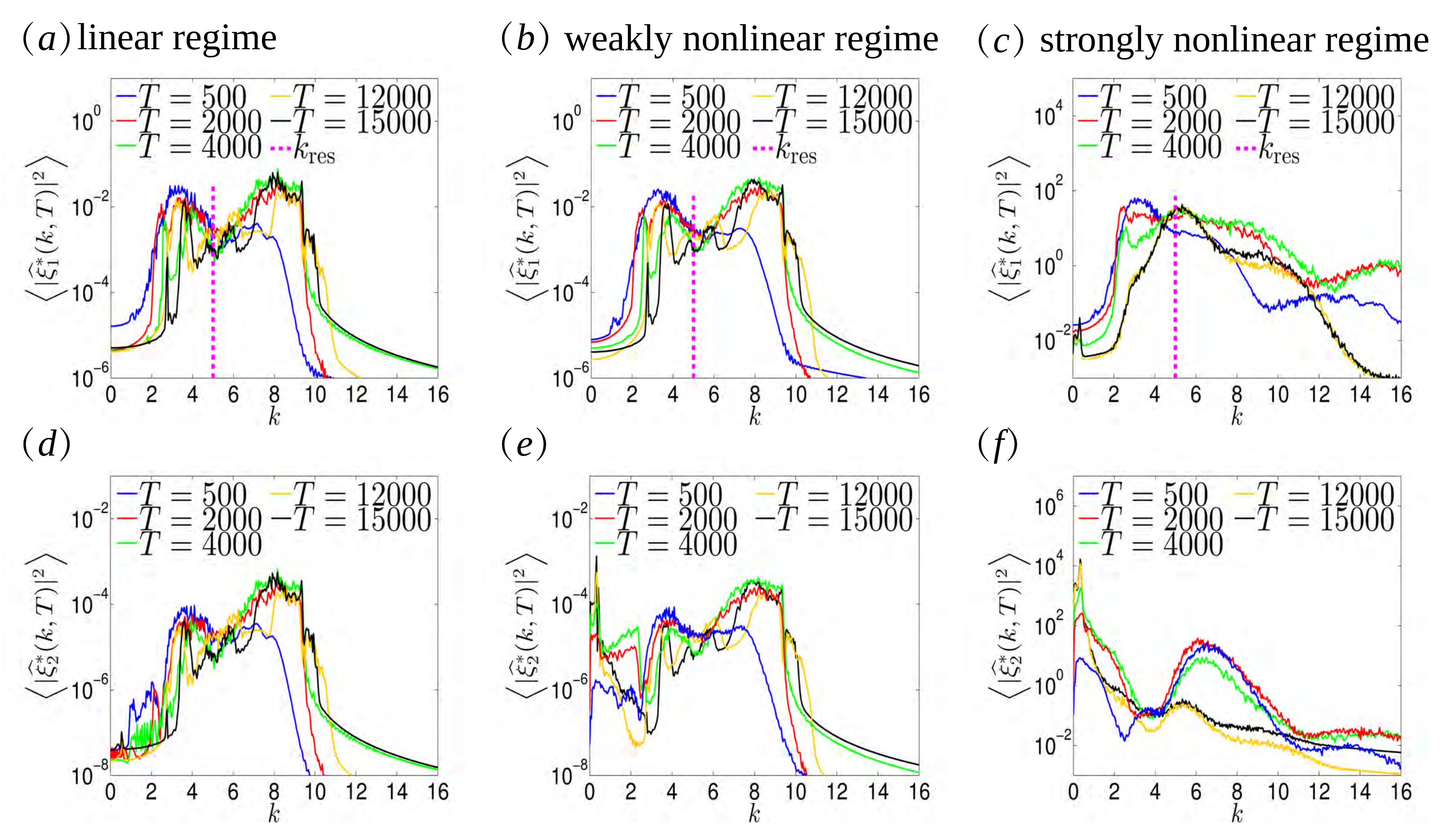}
\caption{(Colour online) Comparison of wavenumber spectra of SWs and IWs in the
linear regime ($\protect\vartheta^{\ast}=0,\protect\sigma_{m}=5$E$-4$) [first
column], in the weakly nonlinear regime ($\protect\vartheta^{\ast}=1,\protect\sigma%
_{m}=5$E$-4$) [second column], and in the strongly nonlinear regime ($%
\protect\vartheta^{\ast}=1,\protect\sigma_{m}=2.5$E$-2$) [third column].
The first row [(\textit{a})-(\textit{c})] displays the
evolution of the wavenumber spectra of SWs' $\protect\xi_{1}^{*} $. The
bracket $\left\langle \cdot \right\rangle $ is interpreted as an ensemble
average over $48$ realizations. The second row [(\textit{d})-(\textit{f})]
displays the evolution of the wavenumber spectra of IWs' $\protect\xi_{2}^{*}$.
All the wavenumber spectra vanish at the initial $T=0$, and these spectra grow as
a result of the external pressure $P_{1}$.
In the linear and weakly nonlinear regimes, the SW spectrum transits between the intervals $[3, 5]$ and $[6.5,9]$ as predicted by the ray-based theory [two green closed curves in figure \ref{Fig:Broad_near_field_theory}(\textit{a})] and thereafter
there is a dip at $k_{\text{res}}$ in the SW spectrum.
In contrast, in the strongly nonlinear regime, there is a peak growing at $k_{\text{res}}$ in the SW spectrum due to the class 3 triad resonance.
Moreover, the IW spectrum grows at $k\approx0.3$ also due to the class 3 triad resonance.
}
\label{Fig:Strong_Spectrum}
\end{figure}

In this section, we will show that in the strongly-nonlinear regime of the mTWN
model, short-mode SWs on the BBF in equation (\ref%
{baroclinic_flow_form}) can be located at the leading edge with the resonant
wavenumber $k_{\text{res}}\approx 5$ [see figure \ref{Fig:class_resonance}(%
\textit{b}) for a description of the resonant wavenumber]. For comparison, we also compute
the short-mode SW behavior in the weakly-nonlinear and linear regimes [see appendix \ref{subsec:limitation}
for more discussion of these regimes]. As introduced previously, $\vartheta^{\ast} =0$ corresponds to the linear regime and $%
\vartheta^{\ast} =1$ corresponds to the nonlinear regime. For $\vartheta^{\ast} =1$, unlike in \S\ \ref{sec:review_lin_non}, we here use
the amplitude of the surface pressure $P_{1}$ to control the strength of the
nonlinearity. Initially, all wave variables $(\xi
_{j}^{\ast },\overline{u}_{j}^{\ast })$ are set to zero. Specifically, we trigger the mTWN system {} by exerting
random pressure $P_{1X}$\ only at the initial $T=0$ as follows:
\begin{equation}
P_{1X}=\sigma _{m}\delta \left( T\right) \sum_{k}\sqrt{\left\vert \tanh %
\left[ s_{L}\left( k-k_{L}\right) \right] -\tanh \left[ s_{R}\left(
k-k_{R}\right) \right] \right\vert }\exp \left( \text{i} kX+\text{i} \phi _{k}\right) ,
\label{initial_force}
\end{equation}%
where $\sigma _{m}$ controls the strength of the initial perturbation, $%
\delta \left( T\right) $ is the Kronecker delta function which is 1 when $%
T=0$ and 0 otherwise, $s_{L}=2.5$, $k_{L}=3$, $s_{R}=1.5$, $k_{R}=4$, and
$\phi _{k}$ is a random phase, which is uniformly distributed on $[0,2\pi ]$
for each wavenumber $k$. Here, the parameters $s_{L}$, $k_{L}$, $s_{R} $,
and $k_{R}$ are empirically chosen such that the initial perturbation waves
possess wavenumbers between the two critical wavenumbers $k_{\text{IW}}$ and $k_{\text{%
res}}$, where $k_{\text{IW}}=0.3$ corresponds to the subsequently-growing
peak location of the IW spectrum [figure \ref{Fig:Strong_Spectrum}] and $k_{%
\text{res}}=5$ is the resonant wavenumber of SWs [figure \ref%
{Fig:Strong_Spectrum}]. Next, we focus on the dynamical behavior of the
short-mode SWs in both the physical $X$ space and the wavenumber $k$ space
for all regimes.

Figure \ref{Fig:Background_Strong} displays the comparison of wave behavior
among the linear, weakly nonlinear, and strongly nonlinear regimes. For the
linear regime ($\vartheta^{\ast} =0$), it can be seen from figures \ref%
{Fig:Background_Strong}(\textit{a})(\textit{d})(\textit{g}) that the SW packets propagate in the
direction of decreasing $X$ with a relatively small group velocity [obvious
dark stripes parallel to the blue line during the time period about $0\leq T\leq 10000$], and in the
direction of increasing $X$ with a relatively large group velocity
[faint dark stripes parallel to the white line during the time period about $10000\leq T\leq 14000$%
]. This zig-zag pattern corresponds to the asymmetric behavior \hyperref[asym1%
]{[1a]}\ of the short-mode SWs, as can be predicted by the linear modulation
theory. In fact, the obvious and faint dark stripes in figures \ref%
{Fig:Background_Strong}(\textit{a})(\textit{g}) reflect the large amplitude and the
small amplitude of the short-mode SWs, respectively. This implies the
asymmetric behavior \hyperref[asym3]{[1c]}, i.e., that SWs become high at
the leading edge and low at the trailing edge.


Moreover, it can be seen from figures \ref%
{Fig:Strong_Spectrum}(\textit{a}) and (\textit{d}) that both SW spectra
and IW spectra are concentrated around the intervals $[3,5]$ and $[6.5,9]$.
Notice that the spectrum of the initial perturbation $P_{1X}$ in equation (\ref%
{initial_force}) is concentrated around the interval $[3,4]$, so that, based
on the linear modulation theory, the wave spectra should be concentrated
around the interval $[6.5,9]$ when SW packets propagate towards the trailing
edge [see the two green-level curves in figure \ref%
{Fig:Broad_near_field_theory}(\textit{a})].
This numerical result implies the asymmetric behavior \hyperref[asym2]{[1b]}%
, i.e., that SWs become short at the leading edge and long at the trailing
edge. Note that no IW spectrum
grows for $k\leq 2$ [figure \ref%
{Fig:Strong_Spectrum}(\textit{d})] and no steep peak forms in the SW spectrum around $k_{%
\text{res}}$ [figure \ref%
{Fig:Strong_Spectrum}(\textit{a})]. Thus, the triad resonance phenomena \hyperref[creso1%
]{[2a]} and \hyperref[creso2]{[2b]} are not observed in this linear regime.
In summary, for the linear regime ($\vartheta^{\ast} =0$), as expected, the
asymmetric behavior \hyperref[asym1]{[1a]}-\hyperref[asym3]{[1c]} is
observed, while the triad resonance phenomena \hyperref[creso1]{[2a]} and
\hyperref[creso2]{[2b]} are not observed. Based on the
ray-based theory, the short-mode waves are linearly modulated by the BBF $(\Xi _{j}^{\ast },\overline{U}_{j}^{\ast })$.




In the weakly nonlinear regime ($\vartheta^{\ast} =1,\sigma _{m}=5$E$-4$), we see
that the spatiotemporal manifestations of SWs and IWs [figures \ref%
{Fig:Background_Strong}(\textit{b})(\textit{e})(\textit{h})] resemble those
of SWs and IWs in the linear regime [figures \ref%
{Fig:Background_Strong}(\textit{a})(\textit{d})(\textit{g})]. In the weakly nonlinear regime, class 3 triad
resonance can occur in the near field.
It can be seen from figure \ref{Fig:Strong_Spectrum}(\textit{b}) that even
the SW spectra resemble those in the linear regime [figure \ref%
{Fig:Strong_Spectrum}(\textit{a})]. The only difference arises from the IW
spectrum. We see from figure \ref{Fig:Strong_Spectrum}(\textit{e}) that
the IW spectrum grows for $k\approx 0.3$ as time increases. This reflects
the triad resonance phenomenon \hyperref[creso1]{[2a]}, i.e., that the
energy is transferred from SWs towards IWs.

For the strongly nonlinear regime ($\vartheta^{\ast} =1,\sigma _{m}=2.5$E$-2$), the
behavior of SWs and IWs, in both the $X$ and $k$ spaces, becomes quite
different from that in the linear and weakly nonlinear regimes. In the
physical $X$ space, surprisingly, one can see from figures \ref%
{Fig:Background_Strong}(\textit{c}) and (\textit{f}) that SW packets are
trapped at the leading edge of the BBF after some
initial time evolution. In the wavenumber $k$ space, both the triad
resonance phenomena \hyperref[creso1]{[2a]} and \hyperref[creso2]{[2b]} are
observed in figure \ref{Fig:Strong_Spectrum}(\textit{f}) and figure \ref%
{Fig:Strong_Spectrum}(\textit{c}), respectively.
It can be seen from figure \ref{Fig:Strong_Spectrum}(\textit{f}) that
IW spectrum grows significantly for $k\approx 0.3$ as time increases, which implies that
a significant amount of energy is
transferred from SWs towards IWs due to the class 3 triad resonance.
It can be seen from figure \ref{Fig:Strong_Spectrum}(\textit{c})
that there is an inverse energy cascade for
the spectrum of SWs from larger wavenumbers above $k_{\text{res}}$
towards smaller wavenumbers, and finally a steep peak forms in the spectrum
of SWs around $k_{\text{res}} $.

We now propose a
possible mechanism, referred to as the modulation-resonance mechanism,
underlying this phenomenon, namely, that SWs whose typical wavenumber is the
resonant wavenumber $k_{\text{res}}$ are located at the leading edge of the
BBF. For convenience of discussing the wave direction, $C_{s}>0$ is assumed
for the velocity of the BBF in the rest of the discussion.

\hypertarget{MECH1}
(\textit{modulation mechanism} [3a]) \label{mr_mech1} First,  after
the initial transients, the large-scale BBF separates
the left-moving and right-moving short-mode waves. The left-moving radiation
quickly leaves the near field, and then only the right-moving waves that
propagate in the same direction as the BBF remain trapped in the near field [figure \ref{Fig:LeftGoing}].


\hypertarget{MECH2}
(\textit{modulation mechanism} [3a']) \label{mr_mech2} Next,  based on the modulation
(ray-based) theory, if the right-moving short-mode perturbation waves
possess wavenumbers below $k_{\text{res}}$, they are transported towards the
leading edge with a relatively large group velocity [long black arrow in
figure \ref{Fig:Broad_near_field_theory}(\textit{a}) and white line
in figure \ref{Fig:Background_Strong}(\textit{i})]. Also, for
right-moving short-mode perturbation waves located in front of the leading
edge with wavenumbers larger than $k_{\text{res}}$ [3 rightmost short black
arrows in figure \ref{Fig:Broad_near_field_theory}(\textit{a})], they are
transported towards the leading edge with a relatively slow group velocity.
Thus, SWs are accumulated at the leading edge. During the sloping process at
the leading edge, these waves become short in wavelength (wavenumber $k$
becomes $k>k_{\text{res}}$) and high in amplitude. Here, the sloping process
means the tendency of short-mode waves to shorten and heighten.

\hypertarget{MECH3}
(\textit{resonance mechanism} [3b]) \label{mr_mech3} Finally, since SW packets
propagate towards the trailing edge with a relatively small group velocity during the sloping process
[3 short black arrows at $X\sim 80$ in figure \ref%
{Fig:Broad_near_field_theory}(\textit{a}) and blue line
in figure \ref{Fig:Background_Strong}(\textit{f})],
energy can be accumulated, and meanwhile the amplitude of SWs grows locally
at the leading edge as a consequence of the asymmetric behavior
\hyperref[asym1]{[1a]}-\hyperref[asym3]{[1c]}.
The growing amplitude increases the
local nonlinearity significantly
and consequently triggers the occurrence of a class 3 triad resonance locally at the
leading edge. Then, an
inverse energy cascade for the spectrum of SWs emerges, streaming from the
larger wavenumbers above $k_{\text{res}}$ towards the smaller wavenumbers,
and finally the spectrum of SWs focuses around the $k_{\text{res}}$
[figure \ref{Fig:Strong_Spectrum}(\textit{c})]. This
spectrum evolution indicates that the class 3 triad resonance condition (\ref%
{triad_bare}) reduces to the resonance condition, $\overline{c}_{g}(k_{\text{%
res}})=C_{s}$, that is, SW packets propagate at the same velocity as the
BBF. Here, $\overline{c}_{g}$ is the modulated group velocity given by $%
\partial \overline{\omega }_{k}/\partial k$. Therefore, SWs with the
resonant wavenumber $k_{\text{res}} $ can be located at the leading edge of
the BBF as shown in figure \ref{Fig:Background_Strong}(\textit{c})
and figure \ref{Fig:Strong_Spectrum}(\textit{c}).

In the following \S\S\  \ref{sec:NESS_setting} and \S\S\ \ref{sec:regime2},
we will examine the robustness of our model results by investigating the
behavior of SWs for various underlying BBFs in a parametric study.
From the parametric study results, we will supplement an additional
\textit{resonance mechanism}
\hyperlink{MECH4}{[3b']}
to the modulation-resonance mechanism in \S\S\ \ref{sec:regime2}.

\subsection{Parametric study for parameter regime I}

\label{sec:NESS_setting}

\begin{table}
\begin{center}
\begin{tabular}{l|c|c|c|c|c|c|c}
Figure & $X_{\Xi ^{\ast }}$ & $s_{\Xi ^{\ast }}$ & $\Xi _{2}^{\ast }(0)$ & $%
\Xi _{1}^{\ast }(0)$ & $\overline{U}_{2}^{\ast }(0)$ & $\overline{U}%
_{1}^{\ast } (0)$ & $C_{s}$ \\[3pt]
\ref{Fig:Background_Strong}(\textit{l}) [red curve] & 120 & ~~0.2 & -0.1 &
0.0003 & ~-0.006 & 0.007 & 0.0485 \\
\ref{Fig:OriginSetup_Wavelength}(\textit{a}) [black curve] & ~20 & ~~0.2 &
-0.1 & 0.0003 & ~-0.006 & 0.007 & 0.0485 \\
\ref{Fig:OriginSetup_Wavelength}(\textit{a}) [blue curve] & ~70 & ~~0.2 &
-0.1 & 0.0003 & ~-0.006 & 0.007 & 0.0485 \\
\ref{Fig:OriginSetup_Wavelength}(\textit{a}) [red curve] & 120 & ~~0.2 & -0.1
& 0.0003 & ~-0.006 & 0.007 & 0.0485 \\
\ref{Fig:OriginSetup_Amplitude}(\textit{e}) [black curve] & ~70 & 0.022 &
-0.1 & 0.0002 & -0.0017 & 0.0044 & 0.0489 \\
\ref{Fig:OriginSetup_Amplitude}(\textit{e}) [blue curve] & ~70 & 0.022 & -0.2
& 0.0004 & -0.0036 & 0.0084 & 0.0502 \\
\ref{Fig:OriginSetup_Amplitude}(\textit{e}) [red curve] & ~70 & 0.022 & -0.3
& 0.0005 & -0.0057 & 0.0120 & 0.0513 \\
&  &  &  &  &  &  &
\end{tabular}%
\end{center}
\caption{The wavelengths $X_{\Xi ^{\ast }}$, slopes $s_{\Xi ^{\ast }}$,
maximal amplitudes of displacements $\Xi _{j}^{\ast }$, maximal amplitudes of currents
$\overline{U}_{j}^{\ast } $, and phase velocities $%
C_{s}$ of the BBF in equation (\protect\ref{baroclinic_flow_form}) used for regime I, $%
(h_{1},h_{2},g,\protect\rho _{1},\protect\rho _{2})=(1,3,1,1,1.003)$. In
figure \protect\ref{Fig:Background_Strong}, the maximal amplitudes of $(\Xi _{j}^{\ast },\overline{U}_{j}^{\ast })$ and
phase velocities
are chosen to be close to those of the interfacial solitary wave
solutions of the TWN model for $\Xi_{2}^{*}(0)=-0.1 $ [see appendix  \protect
\ref{subsec:traveling_wave}]. In figures  \protect
\ref{Fig:OriginSetup_Wavelength} and \protect\ref{Fig:OriginSetup_Amplitude}%
, the maximal amplitudes and phase velocities are chosen to be the same as those of the
interfacial solitary wave solutions of the TWN model [see appendix  \protect
\ref{subsec:traveling_wave}]. }
\label{tab:regime1}
\end{table}

\begin{figure}
\center \hspace*{5mm} 
\includegraphics[scale=1.5]{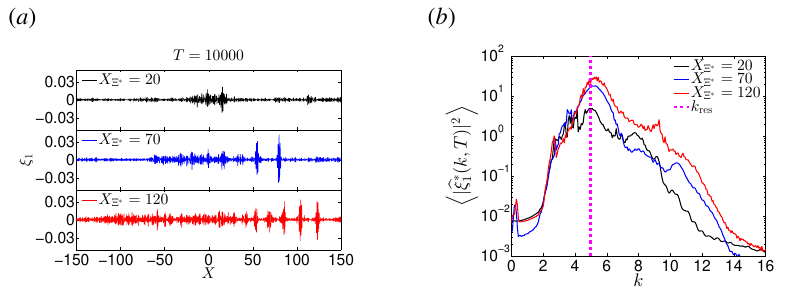}
\caption{(Colour online) Parameter regime I. (\textit{a}) Snapshots of SWs $\protect\xi _{1}$ at
$T=10000$ for different wavelength $X_{\Xi ^{\ast }}$ of the BBF. The parameters
of the BBFs are given in table \ref{tab:regime1}. (\textit{b}%
) The wavenumber spectra of SWs $\protect\xi_{1}^{*} $. The bracket $%
\left\langle \cdot \right\rangle $ is interpreted as the time average over $%
5000 \leq T \leq 10000$.}
\label{Fig:OriginSetup_Wavelength}
\end{figure}

In this section, we investigate how the dynamical behavior of the short-mode
SWs is affected by the wavelength $X_{\Xi ^{\ast }}$, slope $s_{\Xi ^{\ast
}} $, and maximal amplitude of $(\Xi _{j}^{\ast },\overline{U}_{j}^{\ast })$ associated with the
BBF in equation (\ref{baroclinic_flow_form}) [table \ref{tab:regime1}] in the strongly
nonlinear regime of the parameter regime I, $(h_{1},h_{2},g,\rho _{1},\rho
_{2})=(1,3,1,1,1.003)$.
After this parametric study, we will further discuss about the modulation-resonance
mechanism in next \S\S\ \ref{sec:regime2}.
The initial driving force $P_{1X}$\ adopted here is
exactly the same as the force corresponding to the strongly nonlinear regime
in the previous section [equation (\ref{initial_force})]. First, we study the
effects of BBF's wavelength $X_{\Xi ^{\ast }}$ on the dynamical behavior of
short-mode SWs. It can be seen from figure \ref{Fig:OriginSetup_Wavelength}(%
\textit{a}) that the longer the wavelength of the BBF, the larger the number of SW
packets that can be trapped at the leading edge in $X$ space. From figure \ref%
{Fig:OriginSetup_Wavelength}(\textit{b}), it can be seen that the wavenumber
spectra of SWs focus around the resonant wavenumber $k_{\text{res}}$. It can
be further observed that the longer the wavelength of the BBF, the higher
the peak of the SWs' spectrum around the resonant $k_{\text{res}}$.

\begin{figure}
\center \hspace*{5mm} 
\includegraphics[scale=1.3]{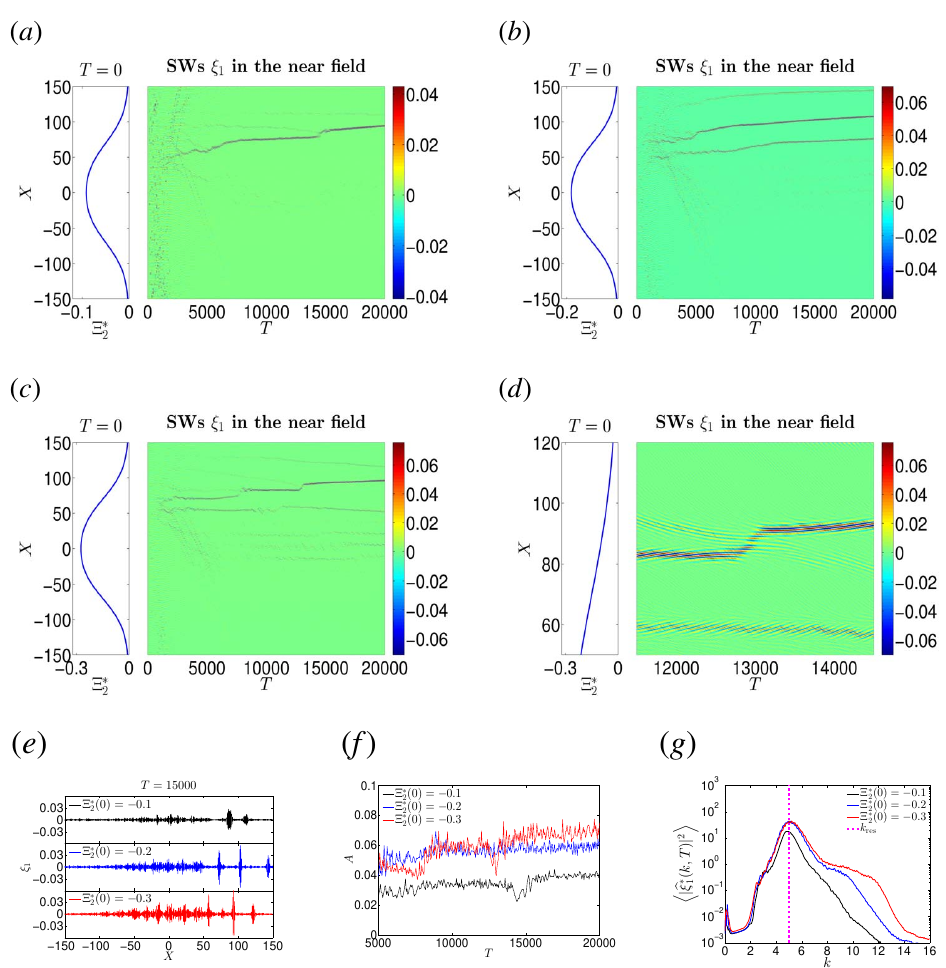}
\caption{(Colour online) Parameter regime I.
The snapshot of the BBF (left panel) and the corresponding spatiotemporal evolution of SWs' profile $%
\protect\xi _{1}$ (right panel) for the different maximal amplitude of the BBF being (\textit{a}) $\Xi_{2}
^{\ast }(0)=-0.1$, (\textit{b}) $\Xi_{2} ^{\ast }(0)=-0.2$, and (\textit{c})
$\Xi_{2} ^{\ast }(0)=-0.3$. The parameters
of the BBFs are given in table \ref{tab:regime1}. (\textit{d}) Zoomed-in version of (\textit{c})
for $11500 \leq T \leq 14500$ for the maximal amplitude of the BBF being $\Xi_{2} ^{\ast
}(0)=-0.3$. (\textit{e}) Snapshots of SWs' profile $\protect\xi _{1}$ at $%
T=15000$ for different maximal amplitudes of the BBFs. (\textit{f}) The temporal
evolution of the maximal amplitude, $A$, of the SWs. (\textit{g}) The
wavenumber spectra of SWs $\protect\xi_{1}^{*} $. The bracket $\left\langle
\cdot \right\rangle $ is interpreted as the time average over $5000 \leq T \leq
20000$. The resonant wavenumber $k_{\text{res}}=5.0$ as depicted by the magenta line. }
\label{Fig:OriginSetup_Amplitude}
\end{figure}

Next, we study the effects of BBF's slope $s_{\Xi ^{\ast }}$ on the
dynamical behavior of short-mode SWs. Note that figures \ref%
{Fig:Background_Strong}--\ref{Fig:OriginSetup_Wavelength} correspond to
the relatively large slope ($s_{\Xi ^{\ast }}=0.2$), whereas figure \ref%
{Fig:OriginSetup_Amplitude} corresponds to the relatively small slope ($%
s_{\Xi ^{\ast }}=0.022$) of the BBF. In $k$ space, wavenumber spectra of SWs
behave similarly, that is, wave spectra focus around the resonant $k_{\text{%
res}}$.
In $X$ space, however, the
spatiotemporal manifestations of SWs differ between large
and small slopes of BBF. For a large slope [figures \ref%
{Fig:Background_Strong} and \ref{Fig:OriginSetup_Wavelength}], SWs are
generated in several size-ordered packets [about 6 packets in figure \ref%
{Fig:Background_Strong}(\textit{l})] with the larger amplitudes appearing at
the leading edge of the BBF, which then slowly decay to smaller amplitudes at the trailing edge.
However, for the small slope, SWs are generated in only one to three
dominant wave packets at the leading edge [figure \ref%
{Fig:OriginSetup_Amplitude}]. For the largest dominant wave packet, both of
its maximal amplitude and group velocity increase after the interaction with
other smaller-amplitude wave packets.
For example, it can be seen from figures \ref{Fig:OriginSetup_Amplitude}(%
\textit{d}) and (\textit{f}) that after the dominant wave packet interacts
with a smaller-amplitude wave packet at $T=13000 $, its amplitude becomes
slightly larger [figure \ref{Fig:OriginSetup_Amplitude}(\textit{f})] and its
group velocity becomes slightly faster [figure \ref%
{Fig:OriginSetup_Amplitude}(\textit{d})]. At the same time, the
small-amplitude wave packet becomes even smaller after the interaction
[figure \ref{Fig:OriginSetup_Amplitude}(\textit{d})]. {During this process,
energy may be transferred from the smaller-amplitude wave packet to the
larger-amplitude one.}

Last, we study the effects of BBF's maximal amplitude, $(\Xi _{j}^{\ast }(0),\overline{U}_{j}^{\ast }(0))$, on the dynamical behavior of
short-mode SWs. It can be seen from figure \ref{Fig:OriginSetup_Amplitude}(%
\textit{f}) that, for larger maximal-amplitude of the BBF, the dominant SW packet
with a larger amplitude can be generated at the leading edge. Also, it can
be seen from figure \ref{Fig:OriginSetup_Amplitude}(\textit{g}) that, for
larger maximal-amplitude of the BBF, the peak of the SWs' spectrum is higher in
amplitude around the resonant $k_{\text{res}}$.

\subsection{Parametric study for parameter regime II}

\label{sec:regime2}

To confirm the modulation-resonance mechanism, we choose a different parameter
regime II, $(h_{1},h_{2},g,\rho _{1},\rho _{2})=(1,5,1,0.856,0.996)$, for
studying the dynamical behavior of the short-mode SWs. This parameter regime
was studied in Ref. \citep{Kodaira2016JFM}. The initial driving force $P_{1X}$\
adopted here is in the form of equation (\ref{initial_force}) with $%
s_{L}=2.5 $, $k_{L}=0.3$, $s_{R}=1.5$, $k_{R}=1$, and $\sigma _{m}=2.5$E$-2$%
. Again, we investigate the effects of the wavelength $X_{\Xi ^{\ast }}$,
slope $s_{\Xi ^{\ast }}$, and maximal amplitude of $(\Xi _{j}^{\ast },\overline{U}%
_{j}^{\ast }) $ associated with the BBF in equation (\ref{baroclinic_flow_form}) [table \ref%
{tab:regime2}] on the short-mode SWs in the strongly nonlinear regime.

\begin{figure}
\center \hspace*{5mm} 
\includegraphics[scale=1.5]{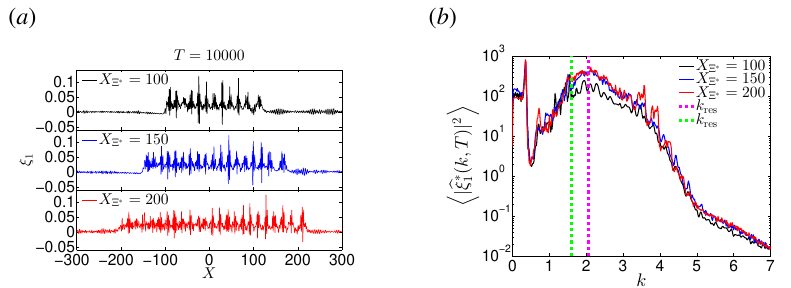}
\caption{(Colour online) Parameter regime II. (\textit{a}) Snapshots of SWs $\protect\xi _{1}$ at
$T=10000$ for different wavelength $X_{\Xi ^{\ast }}$ of the BBF.
The parameters
of the BBFs are given in table \ref{tab:regime2}.
(\textit{b}%
) The wavenumber spectra of SWs $\protect\xi_{1}^{*} $. The bracket $%
\left\langle \cdot \right\rangle $ is interpreted as the time average over $%
5000 \leq T \leq 10000$. The resonant wavenumber $k_{\text{res}}$ depicted by the
green line is $1.6$, where the pure linear dispersion relation $\mu _{k}$ [equation (\ref{Eqn:dispersion_absence})]
is used in the resonance condition (\protect\ref%
{Eqn:resonce_cond}). The resonant wavenumber $k_{\text{res}}$ depicted by the magenta
line is $2.1$, where the modulated dispersion relation (\protect\ref%
{modulated_DR}) with $(\Xi _{1}^{\ast },\Xi _{2}^{\ast },\overline{U}%
_{1}^{\ast },\overline{U}_{2}^{\ast })=(0.025,-0.24,0.08,-0.02) $ is
used in the resonance condition (\protect\ref{Eqn:resonce_cond}).}
\label{Fig:ChoiSetup_Wavelength}
\end{figure}

First, for a relatively small-wavelength BBF ($X_{\Xi ^{\ast }}=100$), SW
packets at the leading edge have almost the same amplitudes as those at the
trailing edge [black curve in figure \ref{Fig:ChoiSetup_Wavelength}(\textit{a%
})]. However, for a relatively large-wavelength BBF ($X_{\Xi ^{\ast }}=150$
and $200$), SWs packets at the leading edge have larger amplitudes
than those at the trailing edge [figure \ref{Fig:ChoiSetup_Wavelength}(%
\textit{a}), blue curve and red curve]. Thus, SW packets can be
observed at the leading edge when the wavelength of the BBF
is sufficiently large. In $k$ space, the spectra of SWs
focus around the resonant wavenumber $k_{\text{res}}$ [magenta line in
figure \ref{Fig:ChoiSetup_Wavelength}(\textit{b})], where the modulated
dispersion relation (\ref{modulated_DR}) [$(\Xi _{1}^{\ast },\Xi _{2}^{\ast
},\overline{U}_{1}^{\ast },\overline{U}_{2}^{\ast
})=(0.025,-0.24,0.08,-0.02) $] is applied in the resonance condition (\ref%
{Eqn:resonce_cond}). For comparison, we also plot the the resonant
wavenumber $k_{\text{res}}$ [green line] where the pure linear dispersion
relation $\mu _{k}$ [equation (\ref{Eqn:dispersion_absence})]
is used in the resonance condition (\ref{Eqn:resonce_cond}). One
can see that the peak locations of the SWs' spectra are in good agreement
with the resonant $k_{\text{res}}$ using the modulated dispersion relation (%
\ref{modulated_DR}) as shown in figure \ref{Fig:ChoiSetup_Wavelength}(\textit{b}). Thus, it is the modulated dispersive waves for
short-modes, instead of the pure linear dispersive waves, that interact with
the long-mode wave through a class 3 triad resonance. [For parameter regime I,
we have not plotted the resonant $k_{\text{res}}$ using the pure linear
dispersion relation for comparison since both resonant wavenumbers $k_{\text{%
res}}$ are very close to each other. That is, the resonant $k_{\text{res}%
}=4.9$ using the pure linear dispersion relation whereas the resonant $k_{%
\text{res}}=5.0$ using the modulated dispersion relation.] Note that it has
been earlier reported by \cite{Kodaira2016JFM} that the modulated dispersion
relation should be used when there is an underlying internal wave.

\begin{table}
\begin{center}
\begin{tabular}{l|c|c|c|c|c|c|c}
Figure & $X_{\Xi ^{\ast }}$ & $s_{\Xi ^{\ast }}$ & $\Xi _{2}^{\ast }(0)$ & $%
\Xi _{1}^{\ast }(0)$ & $\overline{U}_{2}^{\ast }(0)$ & $\overline{U}%
_{1}^{\ast } (0)$ & $C_{s}$ \\[3pt]
\ref{Fig:ChoiSetup_Wavelength}(\textit{a}) [black curve] & 100 & ~~~~0.2 &
-0.24 & 0.025 & -0.02 & 0.08 & 0.38 \\
\ref{Fig:ChoiSetup_Wavelength}(\textit{a}) [blue curve] & 150 & ~~~~0.2 &
-0.24 & 0.025 & -0.02 & 0.08 & 0.38 \\
\ref{Fig:ChoiSetup_Wavelength}(\textit{a}) [red curve] & 200 & ~~~~0.2 &
-0.24 & 0.025 & -0.02 & 0.08 & 0.38 \\
\ref{Fig:ChoiSetup_Amplitude}(\textit{d}) [black curve] & ~90 & 0.0167 &
-0.24 & 0.025 & -0.02 & 0.08 & 0.38 \\
\ref{Fig:ChoiSetup_Amplitude}(\textit{d}) [blue curve] & ~90 & 0.0167 & -0.77
& 0.060 & -0.08 & 0.19 & 0.43 \\
\ref{Fig:ChoiSetup_Amplitude}(\textit{d}) [red curve] & ~90 & 0.0167 & -1.21
& 0.084 & -0.14 & 0.25 & 0.45 \\
&  &  &  &  &  &  &
\end{tabular}%
\end{center}
\caption{The wavelengths $X_{\Xi ^{\ast }}$, slopes $s_{\Xi ^{\ast }}$,
maximal amplitudes $(\Xi _{j}^{\ast }(0),\overline{U}_{j}^{\ast }(0)) $, and velocities $%
C_{s}$ of the BBF in equation (\protect\ref{baroclinic_flow_form}) used for regime II $%
(h_{1},h_{2},g,\protect\rho _{1},\protect\rho _{2})=(1,5,1,0.856,0.996)$
[see this parameter regime in \citep{Kodaira2016JFM}]. In figures \protect\ref%
{Fig:ChoiSetup_Wavelength} and \protect\ref{Fig:ChoiSetup_Amplitude}, the
maximal amplitudes and phase velocities are chosen to be the same as those of the
interfacial solitary wave solutions of the TWN model [see appendix  \protect
\ref{subsec:traveling_wave}].}
\label{tab:regime2}
\end{table}

\begin{figure}
\center \hspace*{5mm} 
\includegraphics[scale=1.03]{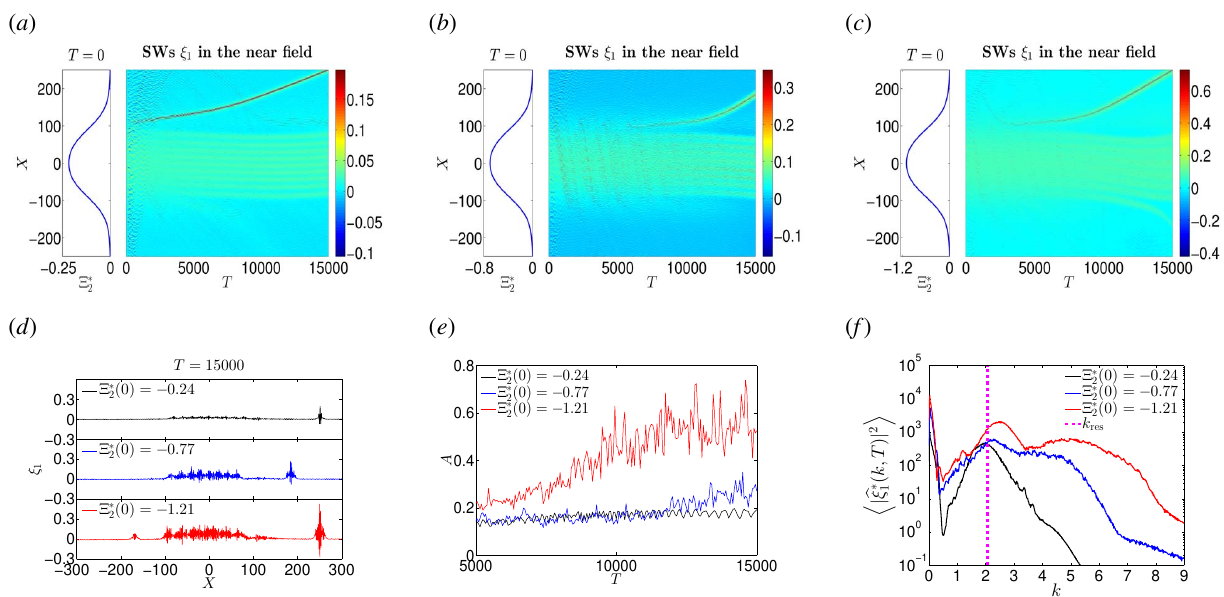}
\caption{(Colour online) Parameter regime II. The snapshot of the BBF (left panel) and the spatiotemporal evolution of SWs' profile $%
\protect\xi _{1}$ (right panel) with the maximal amplitude of the BBF being (\textit{a}) $\Xi_{2}
^{\ast }(0)=-0.24$, (\textit{b}) $\Xi_{2} ^{\ast }(0)=-0.77$, and (\textit{c}%
) $\Xi_{2} ^{\ast }(0)=-1.21$.
The parameters
of the BBFs are given in table \ref{tab:regime2}.
(\textit{d}) Snapshots of SWs' profile $%
\protect\xi _{1}$ at $T=15000$ for different maximal amplitudes of the BBF. (\textit{e%
}) The temporal evolution of the maximal amplitude, $A$, of the SWs. (%
\textit{f}) The wavenumber spectra of SWs' $\protect\xi_{1}^{*} $. The
bracket $\left\langle \cdot \right\rangle $ is interpreted as the time average
over $5000 \leq T \leq 15000$. The resonant wavenumber $k_{\text{res}}=2.1$ as depicted by the magenta line. }
\label{Fig:ChoiSetup_Amplitude}
\end{figure}

Next, by comparing the results in figures \ref{Fig:ChoiSetup_Wavelength} and
\ref{Fig:ChoiSetup_Amplitude}, we can see that SWs are generated in
several wave packets for the large-slope BBF ($s_{\Xi ^{\ast }}=0.2$)
[figure \ref{Fig:ChoiSetup_Wavelength}], whereas SWs are generated in only
one dominant wave packet at the front followed by a small-amplitude
dispersive tail for the small-slope BBF ($s_{\Xi ^{\ast }}=0.0167$) [figure %
\ref{Fig:ChoiSetup_Amplitude}]. Again, the dominant wave packet becomes
higher in amplitude and faster in group velocity after the interaction with
many small-amplitude modulated dispersive waves [figures \ref%
{Fig:ChoiSetup_Amplitude}(\textit{a})(\textit{b})(\textit{c})(e)]. One can
observe that once the amplitude of the dominant wave packet is large
enough, its speed can exceed the speed of the BBF and then move even beyond
the leading edge.


Finally, for the larger-amplitude of the BBF, the dominant SW packet can grow higher
in amplitude at the leading edge [figure \ref{Fig:ChoiSetup_Amplitude}(%
\textit{e})]. In figure \ref%
{Fig:ChoiSetup_Amplitude}(\textit{f}), we still plot the resonant
wavenumber $k_{\text{res}}=2.1$ for reference. One can see that for the
larger-amplitude BBF, the peak of the SW spectrum becomes higher in
amplitude and also the resonant wavenumber becomes slightly larger.

In summary, the effects of BBF's wavelength, slope, and amplitude on the
dynamical behavior of short-mode SWs can be summarized as follows:

\noindent \lbrack 4a] \label{1wvaelength}\textit{SWs can be trapped at the
leading edge only when the wavelength of the BBF is sufficiently long
[figure \ref{Fig:OriginSetup_Wavelength} or figure \ref%
{Fig:ChoiSetup_Wavelength}].}

\noindent \lbrack 4b] \label{2slope}\textit{Fewer numbers of SW packets
are generated at the leading edge when the slope of the BBF is smaller
[comparing figures \ref{Fig:OriginSetup_Wavelength} and \ref%
{Fig:OriginSetup_Amplitude} or comparing figures \ref%
{Fig:ChoiSetup_Wavelength} and \ref{Fig:ChoiSetup_Amplitude}]. }

\noindent \lbrack 4c] \label{3amplitude}\textit{SW packets are larger in
amplitude at the leading edge when the amplitude of the BBF is larger
[figure \ref{Fig:OriginSetup_Amplitude} or figure \ref%
{Fig:ChoiSetup_Amplitude}]. }

Based on the above parametric study of the BBF on the dynamical behavior of short-mode SWs, an additional statement can be supplemented in the description of the modulation-resonance mechanism as discussed in \S\S\ \ref{sec:modu_res}, namely,

\hypertarget{MECH4}
(\textit{resonance mechanism} [3b']) \label{mr_mech4} If
the slope of the
BBF is relatively small, only one or very few dominant ($X$-space localized
large-amplitude) wave packets can be generated at the leading edge. These
localized wave packets become high in amplitude and large in group velocity
after the interaction with other smaller-amplitude wave packets
[figure \ref{Fig:OriginSetup_Amplitude}] or with background perturbation waves [figure \ref%
{Fig:ChoiSetup_Amplitude}].


Our understanding of this
numerical result is provided as follows: Modulation mechanism \hyperlink{MECH2}{[3a']}  cannot
contribute to the growth of the dominant SW packet's amplitude and group
velocity since, in the linear modulation theory, the amplitude of a SW packet
grows only when it propagates towards the trailing edge
[figure \ref{Fig:Broad_near_field_theory}]. Also, dispersive effect cannot
contribute to this growth since dispersion can only reduce the maximal amplitude of the
dominant wave packet [figures \ref{Fig:Broad_near_field_theory} and \ref{Fig:LeftGoing}(\textit{c})]. Thus, only nonlinear effects can contribute to the
growing amplitude of the dominant SW packet. Due to the modulation
mechanism \hyperlink{MECH1}{[3a]}, left-moving and right-moving waves are initially
separated and only right-moving waves remain at the leading edge after
long-time evolution. Therefore, only the nonlinear class 3 triad resonance can
contribute to the growing amplitude of these co-propagating SWs. This
possibly results from the inverse energy cascade flowing from the spectrum of
surrounding perturbation waves with larger wavenumbers above $k_{\text{res}}$\ towards
the dominant SW packets with characteristic wavenumber $k_{\text{res}}$. In
summary, when the slope of the BBF is small, one SW packet or very few SW
packets, localized in both $X$ and $k$ spaces, can be generated at the
leading edge of the underlying BBF.


\section{Conclusions and discussion}

\label{sec:conclusion}

We have investigated the coupling between SWs and IWs on a BBF in our modified two-layer weakly-nonlinear (mTWN) model. The
mTWN model {}\ possesses a class 3 triad resonance between two short-mode
waves and one long-mode wave.
In \S \S\ \ref{sec:linear_modul}, we have reviewed four asymmetric behavior
types of short-mode waves (\hyperref[asym1]{[1a]}-\hyperref[asym4]{[1d]}),
as can be seen in the mTWN model and predicted by
the linear modulation (ray-based) theory. In \S \S\ %
\ref{sec:nonlinear_class3}, using the mTWN model, we have reproduced the class 3 triad resonance
phenomena \hyperref[creso1]{[2a]} and \hyperref[creso2]{[2b]}, as mentioned
by \cite{Tanaka2015JFM}.

\begin{figure}
\centering \hspace*{-10mm} %
\includegraphics[scale=0.37]{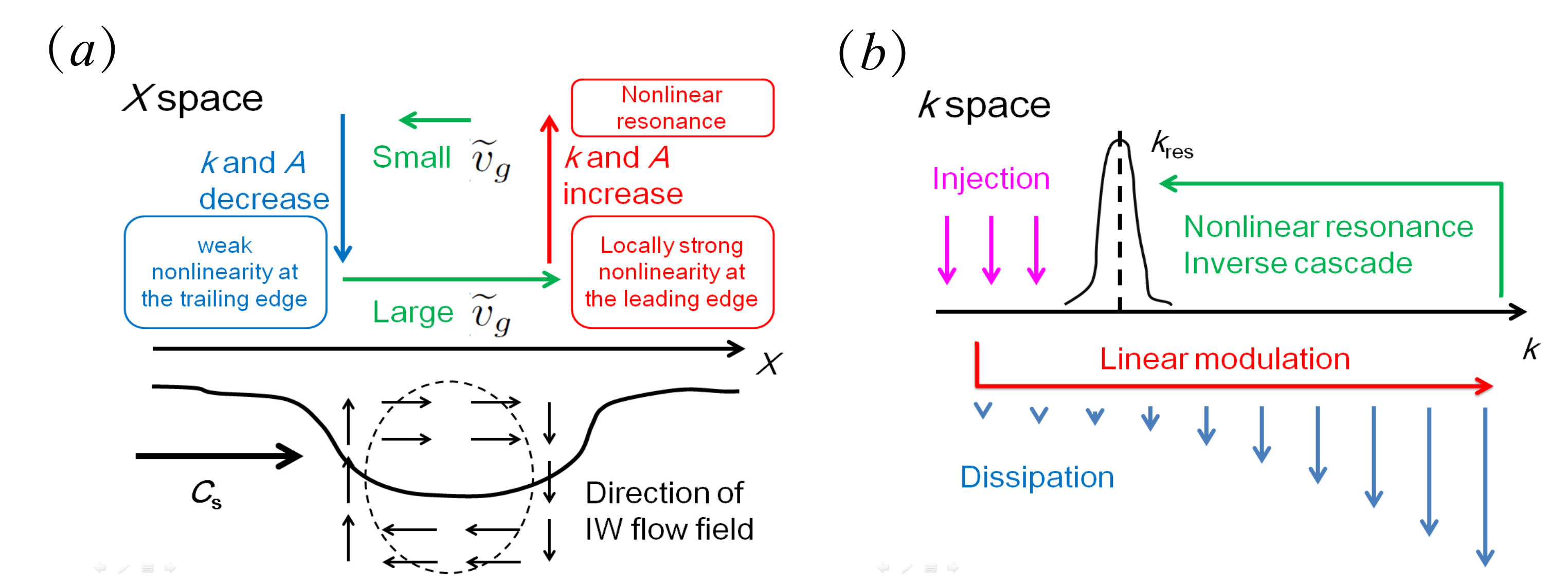}
\caption{(Colour online) Sketch of the modulation-resonance mechanism [see
text] (\textit{a}) in the $X$ space and (\textit{b}) in
the $k$ space. }
\label{Fig:Mechanism_Diagram}
\end{figure}


In \S\ \ref{sec:strong_nonlinear}, we have addressed two important features of SW dynamics driven by
an underlying BBF, that is, a finite number of $X$-space-localized SW packets with $k$-space-localized resonant wavenumber can be generated at the
leading edge of the BBF [figures \ref{Fig:Background_Strong} and \ref{Fig:Strong_Spectrum}].
From our
parametric study, we further observe that, for a sufficiently slowly-varying,
long-wavelength, large-amplitude BBF, one or very few dominant SW packets with growing
amplitude and group velocity can be generated at the leading edge [figures \ref{Fig:OriginSetup_Amplitude} and
\ref{Fig:ChoiSetup_Amplitude}].

Subsequently, we have proposed a possible modulation-resonance mechanism for
understanding these surface phenomena in our numerical results. The
mechanism is sketched in figure \ref{Fig:Mechanism_Diagram}.
First, based on the modulation mechanisms
\hyperlink{MECH1}{[3a]} and \hyperlink{MECH2}{[3a']},
SW packets  propagate with a relatively large group velocity towards the leading edge of the BBF and a sink of SW packets
is thereafter formed at the leading edge.
Next, based on the resonance mechanisms
\hyperlink{MECH3}{[3b]} and \hyperlink{MECH4}{[3b']},
the spectrum of SWs
is concentrated near the resonant wavenumber,
and simultaneously these resonant SWs become high in amplitude at the leading edge.
This modulation-resonance mechanism may be responsible for the experimental observations of
\hyperlink{ASYM_MODU}{(i)}
the surface roughness at the leading edge in the $X$ space and
\hyperlink{CLASS_RESON}{(ii)}
resonant excitations of SWs in the $k$ space \citep{Osborne1980Sci,Kropfli1999JGR}.

As discussed
in the introduction, several previous works used asymptotic
analysis to obtain effective reduced models for addressing
the surface signature of an IW
\citep[][]{Kawahara1975JPSJ,Hashizume1980JPSJp,Funakoshi1983JPSJp,Donato1999JFM,Parau2001JFM,Bakhanov2002JGR,Barros2007SAM,Hwung2009JFM,Craig2012JFM}.
These reduced models mostly focus on either the dynamics of short-mode
waves around the resonant $k_{\text{res}}$ at sufficiently long times or  the linear modulations
and the traveling wave solutions for short-mode surface waves.
A particularly detailed prior discussion of surface signature phenomena was presented in \citep{Craig2012JFM} using
a coupled Korteweg--de Vries (KdV) and linear Schr\"{o}dinger model (we compare this model with ours in appendix \ref{connect}).
However, an intuitive understanding of the surface signature
phenomena was still lacking, especially for the asymmetric behavior of SWs at
the leading and trailing edges.
Our work in the present paper is an attempt to provide some of this understanding.

We now draw several conclusions regarding our work.
First, we point out the reach of our modeling.
The mTWN model consists of Boussinesq-type equations describing the dynamics of
long-wavelength ($10$-$100$ m) short-mode barotropic SWs with an underlying even longer-wavelength
($10$-$1000$ km) baroclinic flow in the shallow-water configuration [see the dimensionless
form of the mTWN model in appendix \ref{subsec:limitation}].
The dispersion relation of the mTWN model differs from that of the Euler system.
The advantage of the mTWN model is in being computationally cheap, which allows us to study
the long-time dynamical behavior of these short-mode SWs.
However, the long-wave approximation (\ref{Eqn:1.assm_beta}) for short-mode SWs is not
appropriate for a quantitative description of  short-wavelength (0.1-1 m) SWs
in the presence of an underlying 1 km internal wave studied in many field observations \citep{Gargettt1972JFM,Osborne1980Sci,Kropfli1999JGR}.
Note that short-mode SWs, corresponding to the dispersion relation $\omega _{k}$ [equation (\ref%
{Eqn:disp_fast})] in the TWN model or the modulated dispersion relation
$\overline{\omega }$ [equation (\ref{modulated_DR})] in the mTWN model,
only means the wavelength of the barotropic SWs is relatively short with respect to the baroclinic flow,
and it does not mean the wavelength of SWs itself is short.
In addition, we point out that our goal in this paper is to only
qualitatively, instead of quantitatively, capture these 0.1-1 m surface signatures of an internal
wave. In future work addressing the quantitative surface signature, the long-wave approximation (\ref{Eqn:1.assm_beta})
should be removed to investigate the dynamical behavior of short-wavelength (0.1-1 m) SWs with an underlying $\sim 1$  km IW in the two-layer Euler system.

Second, we address how our results are
qualitatively consistent with surface phenomena
\hyperlink{ASYM_MODU}{(i)} and \hyperlink{CLASS_RESON}{(ii)}
in field observations. In particular, we point out the following three aspects: {(1) Asymmetric behavior
\hyperref[asym1]{[1a]}-\hyperref[asym4]{[1d]}\ can also be found in the
two-layer Euler system \citep{Lewis1974JFM,Bakhanov2002JGR,Kodaira2016JFM},
and is predicted by the linear modulation (ray-based) theory \citep{Lewis1974JFM,Kodaira2016JFM}. In our
mTWN model, similar asymmetric behavior of SWs can be found and
also predicted by the linear modulation theory [\S \S\ \ref{sec:linear_modul}%
]. (2) The nonlinear class 3 triad resonance phenomena \hyperref[creso1%
]{[2a]} and \hyperref[creso2]{[2b]} were  observed in the two-layer
Euler system by \cite{Tanaka2015JFM}. In our mTWN model, such nonlinear
triad resonance phenomena can also be reproduced [\S \S\ \ref%
{sec:nonlinear_class3}]. }(3) Our model predicts that, in the strongly
nonlinear regime, one spatially-localized large-amplitude SW packet
with wavelength $10$-$100$ m
can be generated at the leading edge of an
underlying slowly-varying baroclinic flow
with wavelength $10$-$1000$ km.
Qualitatively, our result is consistent with the surface phenomena in many
field observations from which a narrow band of $0.1$-$1$ m SWs is located at
the leading edge of an underlying $1$ km internal solitary wave \citep{Gargettt1972JFM,Osborne1980Sci}.
These three aspects allow us to claim that our numerical results are
qualitatively consistent with IW surface signature reported from many field observations.
From the qualitatively-consistent results, we have gained intuition regarding the sizes of two
important parameters $\gamma $ (wavelength ratio of long-mode and short-mode
waves) and $\vartheta $ (nonlinearity) in our mTWN model as discussed in detail in appendix \ref{subsec:limitation}.
It may be helpful to rigorously study the asymptotics of these SWs in the
two-layer Euler system.

Third, our model results predict that, in the weakly nonlinear
regime, short-mode SWs that propagate in the same direction as the
underlying baroclinic flow can be trapped in the near field, propagating back
and forth, and eventually almost uniformly distribute themselves with small amplitudes in the near
field due to the wave dispersion.
This might be related to the phenomenon that small-amplitude SWs 
accompany with  internal tides of the scale around $100$ km \citep{Johnston2003JGR}.
In the strongly nonlinear
regime, our model results predict that very few  SW
packets can be generated locally at the leading edge of the underlying BBF and
they become high in amplitude and large in
group velocity after interacting with the surrounding waves.
The behavior of
these $X$-space-localized, large-amplitude SW packets might be related to
rogue wave phenomena \citep{Muller2005Oceano,Dysthe2008oceanic}.



Finally, also most importantly, we combine the modulation mechanism and resonance mechanism to
understand our numerical results concerning  SWs localized in both $X
$ and $k$ spaces obtained from the mTWN model. We hypothesize that our combined
modulation-resonance mechanism may be extended to the two-layer Euler system in both shallow-
and deep-water configurations and other
density-stratified systems that admit a class 3 triad resonance [see more discussion
in appendix \ref{subsec:limitation}].
As long as
the baroclinic flow has a slight non-uniformity on a sufficiently large
scale, and the perturbation waves are strong enough, a localized wave packet
with the typical resonant wavenumber may be generated at the leading edge of
the baroclinic flow. Further investigation is needed to examine this hypothesis.
We relegate this investigation to future work.

\section{Acknowledgements}

This work is supported by NYU Abu Dhabi Institute G1301(D.C., D.Z.),
NSFC Grant No. 11671259, 11722107, and 91630208, NSFC Grant No. 31571071 (D.C.), and SJTU-UM Collaborative
Research Program (D.C., D.Z.). We thank Yuri V. Lvov and Wooyoung Choi for
helpful comments. We dedicate this paper to our late coauthor and mentor
D.C. who conceived and designed this project.

\appendix

\section{Dimensionless forms and some discussions of the TWN and mTWN
models}

\label{subsec:limitation}

In this appendix, we will provide dimensionless forms of the TWN and mTWN
models and then discuss the asymptotics and typical scaling of the mTWN
model. First, we provide the dimensionless form of the TWN model. Based on
the two approximations $a/h_{1}=\alpha \ll 1$ [equation (\ref%
{Eqn:1.assm_alpha})] and $h_{1}^{2}/l^{2}=\beta \ll 1$ [equation (\ref%
{Eqn:1.assm_beta})], we non-dimensionalize all the variables as follows:%
\begin{align*}
x& =lx^{\prime },\text{ \ \ }t=\left( l/U_{0}\right) t^{\prime }, \\
\overline{u}_{j}& =\alpha U_{0}\overline{u}_{j}^{\prime },\text{ \ }\xi
_{j}=a\xi _{j}^{\prime },\text{ \ }P_{1}=\alpha \rho
_{1}U_{0}^{2}P_{1}^{\prime },
\end{align*}%
where $U_{0}$ is the characteristic speed chosen as $U_{0}=\sqrt{gh_{1}}$, and
all the variables adorned with primes are $O(1)$ in $\alpha $ and $\beta $.
Using asymptotic analysis similar to that for the KdV equation \citep{Whitham1974Linear}
and retaining the terms in the first power of
$\alpha$ and $\beta$, we obtain the TWN model (\ref{Eqn:1.upkin})-(\ref%
{Eqn:1.dndyn}) in the dimensionless form,
\begin{equation}
\eta _{1t^{\prime }}^{\prime }+(\eta _{1}^{\prime }\overline{u}_{1}^{\prime
})_{x^{\prime }}=0,\text{ \ \ }\eta _{1}^{\prime }=h_{1}^{\prime }+\alpha
\xi _{1}^{\prime }-\alpha \xi _{2}^{\prime },  \label{Eqn:1.upkin_less}
\end{equation}%
\begin{equation}
\eta _{2t^{\prime }}^{\prime }+(\eta _{2}^{\prime }\overline{u}_{2}^{\prime
})_{x^{\prime }}=0,\text{ \ \ }\eta _{2}^{\prime }=h_{2}^{\prime }+\alpha
\xi _{2}^{\prime },  \label{Eqn:1.dnkin_less}
\end{equation}%
\begin{equation}
\overline{u}_{1t^{\prime }}^{\prime }+\alpha \overline{u}_{1}^{\prime }%
\overline{u}_{1x^{\prime }}^{\prime }+g^{\prime }\xi _{1x}^{\prime }-\frac{1%
}{3}\beta h_{1}^{\prime 2}\overline{u}_{1x^{\prime }x^{\prime }t^{\prime
}}^{\prime }+\frac{1}{2}\beta h_{1}^{\prime }\xi _{2x^{\prime }t^{\prime
}t^{\prime }}^{\prime }+P_{1x^{\prime }}^{\prime }=O\left( \alpha ^{2},\beta
^{2}\right) ,  \label{Eqn:1.updyn_less}
\end{equation}%
\begin{align}
& \overline{u}_{2t^{\prime }}^{\prime }+\alpha \overline{u}_{2}^{\prime }%
\overline{u}_{2x^{\prime }}^{\prime }+g^{\prime }\xi _{2x^{\prime }}^{\prime
}+\rho _{r}g^{\prime }\left( \xi _{1}^{\prime }-\xi _{2}^{\prime }\right)
_{x^{\prime }}  \label{Eqn:1.dndyn_less} \\
& -\frac{1}{2}\beta \rho _{r}h_{1}^{\prime 2}\overline{u}_{1x^{\prime
}x^{\prime }t^{\prime }}^{\prime }+\beta \rho _{r}h_{1}^{\prime }\xi
_{2x^{\prime }t^{\prime }t^{\prime }}^{\prime }-\frac{1}{3}\beta
h_{2}^{\prime 2}\overline{u}_{2x^{\prime }x^{\prime }t^{\prime }}^{\prime
}+P_{1x^{\prime }}^{\prime }\rho _{1}/\rho _{2}=O\left( \alpha ^{2},\beta
^{2}\right) ,  \notag
\end{align}%
where
\begin{equation*}
h_{1}^{\prime }=1,\text{ \ }h_{2}^{\prime }=h_{2}/h_{1},\text{ \ }g^{\prime
}=1.
\end{equation*}

Next, we provide the derivation of our mTWN model. We begin with the
dimensional TWN model describing the coupling of long waves for two
layers of fluids with unequal densities. The variables $(\xi _{j},\overline{u%
}_{j})$ in the dimensional TWN model are represented as a sum of the BBF
and perturbations of barotropic short-mode waves,
\begin{equation}
(\xi _{j},\overline{u}_{j})=(\Xi _{j}^{\ast },\overline{U}_{j}^{\ast })+(\xi
_{j}^{\ast },\overline{u}_{j}^{\ast }),  \label{Eqn:sum}
\end{equation}%
where $(\Xi _{j}^{\ast },\overline{U}_{j}^{\ast })$\ denotes the BBF and $(\xi
_{j}^{\ast },\overline{u}_{j}^{\ast })$ denotes the perturbation waves. Based on
the assumption (\ref{Eqn:ass_lL}), there is a scale separation of wavelengths between the
BBF $(\Xi _{j}^{\ast },\overline{U}_{j}^{\ast })$ and perturbation waves $%
(\xi _{j}^{\ast },\overline{u}_{j}^{\ast })$. Since we are interested in the
dynamics of short-mode perturbation waves, we choose the characteristic time
and space scalings for these short-mode waves such that the chosen scalings
are self-consistent with our numerical results for the TWN model. We
non-dimensionalize all variables $(\Xi _{j}^{\ast },\overline{U}_{j}^{\ast
}) $ and $(\xi _{j}^{\ast },\overline{u}_{j}^{\ast })$\ as follows:

\begin{align}
x& =lx^{\prime },\text{ \ \ }t=\left( l/U_{0}\right)
t^{\prime },\text{ \ }P_{1}=\rho _{1}u_{0}U_{0}P_{1}^{\prime },
\notag \\
\overline{U}_{j}^{\ast }& =U_{0}\overline{U}_{j}^{\prime },\text{ \ }%
\overline{u}_{j}^{\ast }=u_{0}\overline{u}_{j}^{\prime },\text{ \ }\Xi
_{j}^{\ast }=A\Xi _{j}^{\prime },\text{ \ }\xi _{j}^{\ast }=a\xi
_{j}^{\prime },  \label{Eqn:scaling}
\end{align}%
where $l$ is the characteristic wavelength of the short-mode waves, $U_{0}$
and $u_{0}$\ are the characteristic speeds of the BBF and the short-mode
waves, respectively, and $A$ and $a$ are the characteristic amplitudes of
the BBF and the short-mode waves, respectively. We further apply the
following assumptions:

[1] The dimensionless BBF, $%
(\Xi _{j}^{\prime },\overline{U}_{j}^{\prime })$, is a right-moving
traveling wave with the phase velocity $C_{s}$ satisfying,
\begin{equation}
\Xi _{j}^{\prime }=\Xi _{j}^{\prime }\left( X^{\prime }\right) ,\text{ \ }%
\overline{U}_{j}^{\prime }=\overline{U}_{j}^{\prime }\left( X^{\prime
}\right) \text{ \ with }X^{\prime }=K\left( x-C_{s}t\right) ,\text{ }K=2\pi
/L,\text{ \ }  \label{Eqn:ass_travel2}
\end{equation}%
where $L$ is the characteristic wavelength of the BBF. Here, the derivatives
of the dimensionless BBF are of $O\left( 1\right) $, that is, $d\Xi
_{j}^{\prime }/dX^{\prime }=O\left( 1\right) $ and $d\overline{U}%
_{j}^{\prime }/dX^{\prime }=O\left( 1\right) $.

[2] The characteristic wavelength of the short-mode
waves, $l$, is much shorter than the characteristic wavelength of the BBF, $%
L $, \citep{Hwung2009JFM}
\begin{equation}
2\pi l/L=\gamma \ll 1.  \label{Eqn:ass4}
\end{equation}%

[3] For the BBF, the characteristic
amplitude $A$ and the characteristic speed $U_{0}$ satisfy,
\begin{equation}
A=h_{1},\text{ \ \ }U_{0}=\sqrt{gh_{1}}.  \label{Eqn:ass3}
\end{equation}%

[4] For the short-mode waves, the characteristic amplitude
$a$ and the characteristic speed $u_{0}$ satisfy
\begin{equation}
a=\vartheta h_{1},\text{ \ \ }u_{0}=\vartheta \sqrt{gh_{1}},\text{ \ }
\label{Eqn:ass2}
\end{equation}%
where the parameter $\vartheta $\ controls the relative scaling between the short-mode waves and the
BBF.

[5] The characteristic wavelength of the
short-mode wave, $l$, satisfies
\begin{equation}
l=h_{1},  \label{Eqn:ass1}
\end{equation}%
which is self-consistent with our numerical results [figures \ref%
{Fig:bare_dispersion}-\ref{Fig:ChoiSetup_Amplitude}]. Here, note that from
the numerical results, {the characteristic wavelength of the resonant
short-mode waves is }$l\sim 2\pi /k_{\text{res}}=1.25$, which is on the
same order of $h_{1}=1$ [see figure \ref{Fig:bare_dispersion}(\textit{b})
for example].

All variables adorned with primes are $O(1)$ in $\gamma $. Substituting the
sum (\ref{Eqn:sum}), the non-dimensionalization (\ref{Eqn:scaling}), and the
assumptions (\ref{Eqn:ass_travel2})-(\ref{Eqn:ass1}) into our TWN model, we obtain
the model in the following dimensionless form:%
\begin{align}
& \eta _{1t^{\prime }}^{\prime }+\left[ h_{1}^{\prime }\overline{u}%
_{1}^{\prime }+\left( \Xi _{1}^{\prime }-\Xi _{2}^{\prime }\right) \overline{%
u}_{1}^{\prime }+\overline{U}_{1}^{\prime }(\xi _{1}^{\prime }-\xi
_{2}^{\prime })+\vartheta (\xi _{1}^{\prime }-\xi _{2}^{\prime })\overline{u}%
_{1}^{\prime }\right] _{x^{\prime }}  \label{Eqn:effmtwn_1} \\
& -\frac{\gamma }{\vartheta }\frac{C_{s}}{U_{0}}H_{1X^{\prime }}^{\prime }+%
\frac{\gamma }{\vartheta }h_{1}^{\prime }\overline{U}_{1X^{\prime }}^{\prime
}+\frac{\gamma }{\vartheta }\left( \left( \Xi _{1}^{\prime }-\Xi
_{2}^{\prime }\right) \overline{U}_{1}^{\prime }\right) _{X^{\prime }}=0,
\notag
\end{align}%
\begin{align}
& \eta _{2t^{\prime }}^{\prime }+\left[ h_{2}^{\prime }\overline{u}%
_{2}^{\prime }+\Xi _{2}^{\prime }\overline{u}_{2}^{\prime }+\overline{U}%
_{2}^{\prime }\xi _{2}^{\prime }+\vartheta \xi _{2}^{\prime }\overline{u}%
_{2}^{\prime }\right] _{x^{\prime }}  \label{Eqn:effmtwn_2} \\
& -\frac{\gamma }{\vartheta }\frac{C_{s}}{U_{0}}H_{2X^{\prime }}^{\prime }+%
\frac{\gamma }{\vartheta }h_{2}^{\prime }\overline{U}_{2X^{\prime }}^{\prime
}+\frac{\gamma }{\vartheta }\left( \overline{U}_{2}^{\prime }\Xi
_{2}^{\prime }\right) _{X^{\prime }}=0,  \notag
\end{align}%
\begin{align}
& \overline{u}_{1t^{\prime }}^{\prime }+\left( \overline{U}_{1}^{\prime }%
\overline{u}_{1}^{\prime }+\frac{1}{2}\vartheta \overline{u}_{1}^{\prime
2}+g^{\prime }\xi _{1}^{\prime }\right) _{x^{\prime }}-\frac{1}{3}%
h_{1}^{\prime 2}\overline{u}_{1x^{\prime }x^{\prime }t^{\prime }}^{\prime }+%
\frac{1}{2}h_{1}^{\prime }\xi _{2x^{\prime }t^{\prime }t^{\prime }}^{\prime
}+P_{1x^{\prime }}^{\prime }  \label{Eqn:effmtwn_3} \\
& -\frac{\gamma }{\vartheta }\frac{C_{s}}{U_{0}}\overline{U}_{1X^{\prime
}}^{\prime }+\frac{\gamma }{\vartheta }\overline{U}_{1}^{\prime }\overline{U}%
_{1X^{\prime }}^{\prime }+\frac{\gamma }{\vartheta }g^{\prime }\Xi
_{1X^{\prime }}^{\prime }+\frac{1}{3}\frac{\gamma ^{3}}{\vartheta }\frac{%
C_{s}}{U_{0}}h_{1}^{\prime 2}\overline{U}_{1X^{\prime }X^{\prime }X^{\prime
}}^{\prime }+\frac{1}{2}\frac{\gamma ^{3}}{\vartheta }\frac{C_{s}^{2}}{%
U_{0}^{2}}h_{1}^{\prime }\Xi _{2X^{\prime }X^{\prime }X^{\prime }}^{\prime
}=0,  \notag
\end{align}%
\begin{align}
& \overline{u}_{2t^{\prime }}^{\prime }+\left( \overline{U}_{2}^{\prime }%
\overline{u}_{2}^{\prime }+\frac{1}{2}\vartheta \overline{u}_{2}^{\prime
2}+g^{\prime }\xi _{2}^{\prime }+\rho _{r}g^{\prime }\eta _{1}^{\prime
}\right) _{x^{\prime }}  \label{Eqn:effmtwn_4} \\
& -\frac{1}{2}\rho _{r}h_{1}^{\prime 2}\overline{u}_{1x^{\prime }x^{\prime
}t^{\prime }}^{\prime }+\rho _{r}h_{1}^{\prime }\xi _{2x^{\prime }t^{\prime
}t^{\prime }}^{\prime }-\frac{1}{3}h_{2}^{\prime 2}\overline{u}_{2x^{\prime
}x^{\prime }t^{\prime }}^{\prime }+P_{1x^{\prime }}^{\prime }\rho _{1}/\rho
_{2}  \notag \\
& -\frac{\gamma }{\vartheta }\frac{C_{s}}{U_{0}}\overline{U}_{2X^{\prime
}}^{\prime }+\frac{\gamma }{\vartheta }\overline{U}_{2}^{\prime }\overline{U}%
_{2X^{\prime }}^{\prime }+\frac{\gamma }{\vartheta }g^{\prime }\Xi
_{2X^{\prime }}^{\prime }+\frac{\gamma }{\vartheta }\rho _{r}g^{\prime
}H_{1X^{\prime }}^{\prime }  \notag \\
& +\frac{1}{2}\frac{\gamma ^{3}}{\vartheta }\frac{C_{s}}{U_{0}}\rho
_{r}h_{1}^{\prime 2}\overline{U}_{1X^{\prime }X^{\prime }X^{\prime
}}^{\prime }+\frac{\gamma ^{3}}{\vartheta }\frac{C_{s}^{2}}{U_{0}^{2}}\rho
_{r}h_{1}^{\prime }\Xi _{2X^{\prime }X^{\prime }X^{\prime }}^{\prime }+\frac{%
1}{3}\frac{\gamma ^{3}}{\vartheta }\frac{C_{s}}{U_{0}}h_{2}^{\prime 2}%
\overline{u}_{2X^{\prime }X^{\prime }X^{\prime }}^{\prime }=0,  \notag
\end{align}%
where%
\begin{equation*}
h_{1}^{\prime }=1,\text{ \ }h_{2}^{\prime }=h_{2}/h_{1},\text{ \ }g^{\prime
}=1,
\end{equation*}%
\begin{equation*}
H_{1}^{\prime }=h_{1}^{\prime }+\Xi _{1}^{\prime }-\Xi _{2}^{\prime },\text{
\ \ }H_{2}^{\prime }=h_{2}^{\prime }+\Xi _{2}^{\prime },\text{ \ }
\end{equation*}%
\begin{equation*}
\eta _{1}^{\prime }=h_{1}^{\prime }+\xi _{1}^{\prime }-\xi _{2}^{\prime },%
\text{ \ \ }\eta _{2}^{\prime }=h_{2}^{\prime }+\xi _{2}^{\prime }.
\end{equation*}%
In the limit of
\begin{equation}
\gamma \rightarrow 0,  \label{Eqn:ass5}
\end{equation}%
we obtain the dimensionless mTWN model. Casting the mTWN system in the conservation
form in the right-moving frame with the velocity $C_{s}$, we can obtain our
mTWN model, in dimensional form, in the equations of (\ref{Eqn:eff_1})-(\ref{Eqn:eff_4}).
Since the interfacial solitary wave solutions in the TWN model [see figure \ref{Fig:broaden}
in the following appendix \ref{subsec:traveling_wave}] are not long
enough in wavelength to satisfy the assumption (\ref{Eqn:ass5}), we instead use a slowly-varying, long-wavelength BBF  in the mTWN model.
This is the
reason why we use the mTWN model instead of the TWN model in our investigation.

We now discuss the asymptotics for the mTWN model.

\noindent [\textit{1}] If $%
\vartheta =0$ (or equivalently $\vartheta \rightarrow 0$ and $\gamma / \vartheta \rightarrow 0$), the model corresponds to the linear regime [1st column of
figure \ref{Fig:Background_Strong}];

\noindent [\textit{2}] if $\vartheta \ll 1$, the
model corresponds to the weakly nonlinear regime [2nd column of figure \ref%
{Fig:Background_Strong}];

\noindent [\textit{3}] if $\vartheta =O\left( 1\right) $,
the model corresponds to the strongly nonlinear regime [3rd column of figure %
\ref{Fig:Background_Strong}].

When $\vartheta = 0$,
short-mode waves satisfy the modulated dispersion relation (\ref%
{modulated_DR}) instead of the pure linear dispersion relation (\ref%
{Eqn:dispersion_absence}). For short-time dynamics, we can see the phase
velocity of the short-mode waves [red lines in figures \ref{Fig:Background_Strong}(%
\textit{d})-(\textit{f})]. When $\vartheta  $ is nonzero, dispersive
waves are coupled through the class 3 triad resonance. For long-time dynamics,
we can see the group velocity of short-mode waves [blue and white lines in figures \ref%
{Fig:Background_Strong}(\textit{a})-(\textit{c})].
Here, for asymptotic analysis, we suggest using the modulated dispersion relation
instead of the pure linear dispersion relation at the leading order as in \citep{Kodaira2016JFM}.

Next, we turn to the discussion of the sizes of the parameters. There are two
important parameters in the mTWN model, one being $\vartheta $ controlling the
strength of the nonlinearity, and the other being $\gamma $ describing the
wavelength ratio between the short-mode waves and the BBF. The effects of the
nonlinearity strength $\vartheta $ can be seen from the comparisons among
the linear, weakly-nonlinear, and strongly-nonlinear regimes in figure \ref%
{Fig:Background_Strong}. The effects of the parameter $\gamma $ can be seen
from the parametric study in \S \S\ \ref{sec:NESS_setting} and \S \S\ \ref%
{sec:regime2}. {The scale separation} between the BBF and short-mode waves
is consistent with the assumption of the application of the linear
modulation theory and the class 3 triad resonance theory. For{\ the linear
modulation theory, the wavelength of the BBF needs to be sufficiently long
so that the BBF can be nearly treated as constant as compared to the short-mode waves. For the
triad resonance theory, the condition, }$\Delta k\ll k_{\text{res}}$,{\ is
needed to ensure the occurrence of a triad resonance among one long-mode wave
and two short-mode waves}. In fact, we can also begin with the Euler system
to obtain the mTWN model assuming the scaling relation $\gamma \ll \alpha
,\beta \ll 1$, where $\alpha $ in equation (\ref{Eqn:1.assm_alpha}) and $\beta $ in equation (%
\ref{Eqn:1.assm_beta}) are assumed only for short-mode waves. From the above
discussion, we provide intuition concerning the size of these parameters from
which one can understand surface phenomena
\hyperlink{ASYM_MODU}{(i)} and \hyperlink{CLASS_RESON}{(ii)}.

Finally, we briefly discuss the BBF in the mTWN model. Physically,
if the length of the short-mode waves is long
[about 10-100 m], then the BBF corresponds to a
baroclinic flow across the ocean [about $10$-$1000$ km or even longer] based
on the assumption (\ref{Eqn:ass4}).
For the dynamics of such
long-wavelegnth baroclinic flows, the effects of internal tides, Coriolis force, variable topography, and continuous
density-stratification become important, which are not involved in this work.
For example, internal tides, internal waves at a tidal frequency, can have wavelength about 150 km
\citep{Johnston2003JGR}.
The BBF  may be thereafter regarded as an external baroclinic flow induced by a sufficiently long-wavelength
IW, such as internal tides.

\section{Interfacial solitary wave solutions}

\label{subsec:traveling_wave}

\begin{figure}
\centering  \hspace*{-2.0mm} %
\includegraphics[width=5.3in,height=2.1in]{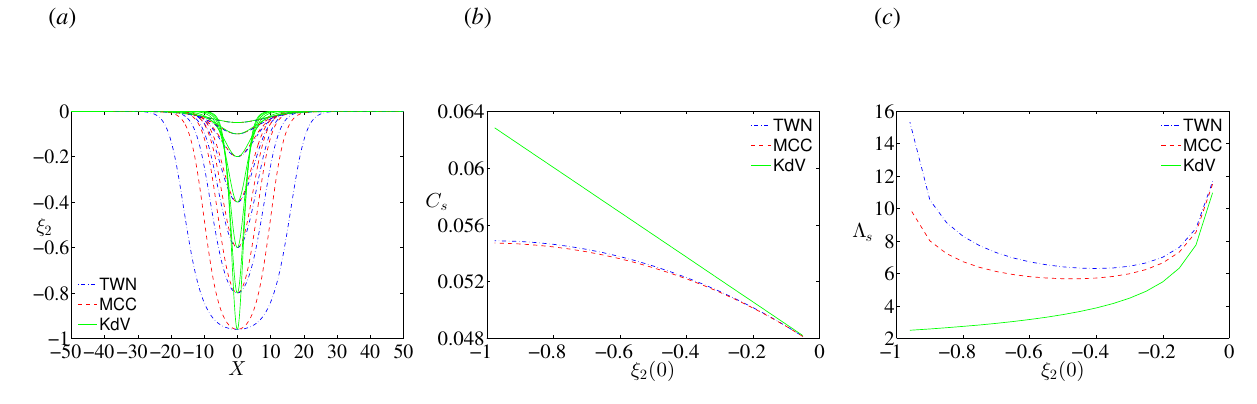}
\caption{(Colour online) (\textit{a}) Interfacial
solitary-wave solutions of the TWN model (blue dashed-dotted line
--\thinspace $\cdot $\thinspace --\thinspace $\cdot $\thinspace --), the MCC
model (red dashed lines --\, --\, --) \citep{Choi1999JFM},
and the KdV model (green solid lines ---------)
\citep{Choi1999JFM} with the same amplitude. As the amplitude increases, the
solitary-wave solutions of both TWN and MCC models broaden and eventually develop a flat crest when the
amplitude increases to approximately $(h_{1}-h_{2})/2$. (\textit{b})
Velocity $C_{s}$ versus wave amplitude $\protect\xi_{2}(0) $. There is good
agreement of velocity between the TWN model and MCC model. (\textit{c})
Effective width $\Lambda_{s}$ versus wave amplitude $\protect\xi_{2}(0)$. There is
qualitatively good agreement of effective width between the TWN model and MCC model.}
\label{Fig:broaden}
\end{figure}

In this appendix, we present the interfacial solitary wave solutions to the
TWN system.
We numerically
compute the right-moving interfacial solitary wave solutions that propagate
with the velocity $C_{s}$ using the numerical methods of \cite{Han2007AMC}.
Figure \ref{Fig:broaden}(\textit{a}) displays solitary wave solutions of the
TWN system for IWs with different amplitudes. We also plot
the corresponding MCC and KdV solutions with the same amplitudes %
\citep{Choi1999JFM}. It can be seen from figure \ref{Fig:broaden}(\textit{a}%
) that the TWN model can capture the broadening of IWs which is often
observed in the ocean \citep{Duda2004IEEEJOE}. Note that the broadening of
interfacial solitary wave solutions can also be captured by other models %
\citep{Craig2005CPAM,Guyenne2006CRM}, not only by the MCC model. For example,
\cite{Guyenne2006CRM} proposed a Hamiltonian model of a two-layer
stratification whose long internal wave solutions can well capture the
broadening of strongly nonlinear solitary waves. Here, we reproduce this
property for the interfacial solitary wave using the TWN model. The maximum
amplitude of IWs in the TWN model approximately equals $(h_{1}-h_{2})/2$.
Figure \ref{Fig:broaden}(\textit{b}) displays the comparison of the velocity
$C_{s}$ among these three models. It can be seen that the velocity of the
TWN model is in good agreement with that of the MCC model, whereas it
deviates greatly from the KdV model for large values of the wave
amplitude $\xi _{2}(0)$.

To quantify the trend of wavelength broadening, we introduce a measure of
the effective width, $\Lambda _{s}$, for the interfacial solitary wave
solution \citep{Koop1981JFM}:
\begin{equation}
\Lambda _{s}=\left\vert \frac{1}{\xi _{2}(0)}\int_{0}^{\infty }\xi
_{2}(X)dX\right\vert .  \label{Eqn:soliton_speed}
\end{equation}%
For the effective width of the MCC and KdV solutions, one can refer to the
reference \citep{Choi1999JFM}. Figure \ref{Fig:broaden}(\textit{c}) displays
the effective widths among TWN solutions, MCC solutions, and
KdV solutions. For small-amplitude waves, there is good agreement of
effective widths among the three models. However, discrepancies grow rapidly
with the increase of the wave amplitude. {Note that in this work, we focus
more on the qualitative behavior of SWs instead of IWs. For IWs, the
key point here is that our TWN model, MCC-type model, and the Hamiltonian model
by \cite{Guyenne2006CRM} can capture the broadening of IWs as can be seen in
the ocean \citep{Duda2004IEEEJOE} while the KdV model cannot.}

\section{Numerical method and low-pass filter}

\label{accuracy_test}

Here, we present some details pertaining to our numerical computation.
Throughout the text, the variables are all dimensionless. For the parameter
regime I [$(h_{1},h_{2},g,\rho _{1},\rho _{2})=(1,3,1,1,1.003)$], we use $%
2^{15}$ uniformly distributed grid points for the numerical examples. We set
the computational domain to be $[-M,M]$, with $M=300$. At the boundaries $%
X=-M $ and $X=M$, we use periodic boundary conditions. For the numerical
examples in \S \S\ \ref{sec:linear_modul} and \S\ \ref{sec:strong_nonlinear}%
, we find that even for an initially narrowly localized short-mode wave
perturbation, radiation is quickly emitted from the near field
towards the boundaries $X=-M$ and $X=M$. To eliminate possible reflected
waves from these boundaries, we establish two buffer zones in the regions $%
[-M,-M+m]$ and $[M-m,M]$ ($m\ll M$) where we add damping and diffusion terms
to absorb the outgoing radiation. For the parameter regime II [$%
(h_{1},h_{2},g,\rho _{1},\rho _{2})=(1,5,1,0.856,0.996)$], we use $5\times
2^{13}$ grid points and domain with $M=375$. For our numerical scheme, we
use the fourth-order Runge-Kutta method in time and the collocation method
in space \citep{Chen2005Efficient}. We have numerically examined the
accuracy of this scheme for the interfacial solitary wave solution. We use
the $2/3$-rule for de-aliasing and a low-pass filter to suppress the Kelvin-Helmholtz instability
as discussed below.

\begin{figure}
\centering 
\includegraphics[width=5.1in]{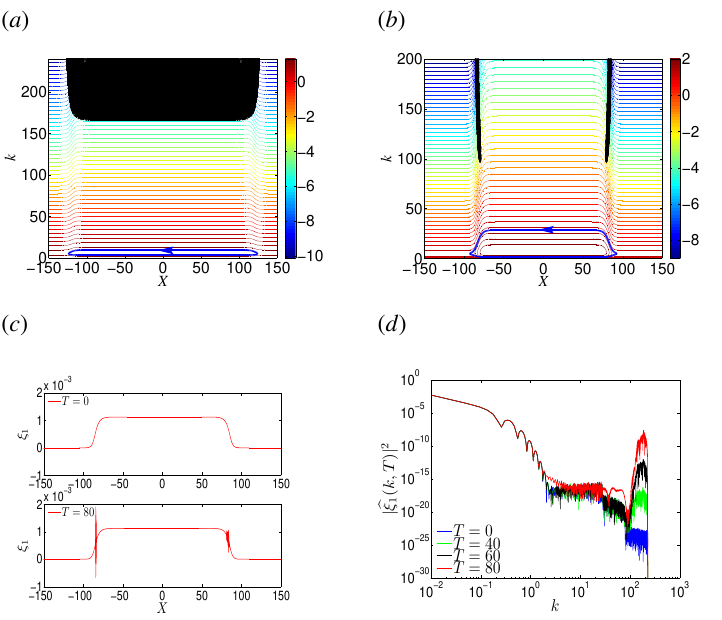}
\caption{(Colour online) (\textit{a}) The phase portrait of the short-mode
wave-packet motion on the BBF in equation (\protect\ref%
{baroclinic_flow_form}). Along each curve, the frequency $\overline{\protect%
\nu }_{k}$ remains constant, with its value color-coded. Inside the region
enclosed by the blue level-curve, the short-mode waves are trapped. The
arrows indicate the direction of the wave-packet movement. Inside the black
region, the frequency $\overline{\protect\nu }_{k}$ is not real-valued,
\textit{i.e.}, the mTWN model {} suffers from the KH instability in this
region. (\textit{b}) The phase portrait of the short-mode wave-packet motion
corresponding to an underlying interfacial solitary wave with the amplitude $%
0.976$. Inside the two black regions, the mTWN model suffers from the KH
instability. (\textit{c}) Snapshots of SWs' profile $\protect\xi_{1}$ at $%
T=0 $ and $T=80$, when the low-pass filter is not applied in the simulation.
(\textit{d}) Time evolution of SWs' wavenumber spectrum $|\widehat{\protect%
\xi }_{1}(k,T )|^{2} $ when the low-pass filter is not applied in the
simulation. }
\label{Fig:KH_Analysis}
\end{figure}

Next, we demonstrate that although our mTWN system suffers from the
Kelvin-Helmholtz (KH) instability in the presence of the BBF (\ref{baroclinic_flow_form}), these unstable wavenumbers are
relatively large and are outside the wavenumber domain in some of our examples.
Hence, our numerical computation is still stable and does not suffer from
the KH instability. Figure \ref{Fig:KH_Analysis}(\textit{a}) reproduces the
phase portrait of the wave-packet motion depicted in figure \ref%
{Fig:Broad_near_field_theory}(\textit{a}), but over a larger wavenumber $k$\
range. In the black region, the frequency $\overline{\nu }_{k}$ is not
real-valued so that our mTWN system suffers from the KH instability. One can
observe that the KH instability occurs for wavenumbers $k > k_{%
\text{KH}}=170$. However, this critical wavenumber $k_{\text{KH}}$ is
outside the wavenumber domain used in our computation. Thus, our computation is
still numerically stable.


However, for large-amplitude interfacial solitary wave simulations, our
numerical computation suffers from the KH instability. Figure \ref%
{Fig:KH_Analysis}(\textit{b}) displays the phase portrait of the wave-packet
motion on an interfacial solitary wave with the amplitude $0.976$. We can
see that the KH instability occurs at the leading and trailing edges for
wavenumbers $k$ greater than $k_{\text{KH}}=100$. We numerically simulate
our model using $2^{16}$ uniformly spaced grid points. The results are shown
in figures \ref{Fig:KH_Analysis}(\textit{c}) and (\textit{d}): the KH
instability, as expected, occurs at the leading and trailing edges for
wavenumbers $k$ greater than $k_{\text{KH}}=100$.

To suppress this KH instability, we adopt in our computation a low-pass
filter as in \citep{Jo2008SAM}. In particular, we inhibit the growth of
waves with their wavenumbers greater than $k_{\text{KH}}=90$. To examine the
validity of the low-pass filter, we compute the conservation of mass and
momentum,%
\begin{equation}
C_{\xi _{j}}(T)=\int \xi _{j}dX,\ \ \ C_{M_{j}}(T)=\int M_{j}dX,
\label{conservation_law}
\end{equation}%
and then measure the maximum relative errors,%
\begin{equation}
\Delta C_{\xi _{j}}=\max_{T}\left\vert \frac{C_{\xi _{j}}(T)-C_{\xi _{j}}(0)%
}{C_{\xi _{j}}(0)}\right\vert ,\text{ \ }\Delta C_{M_{j}}=\max_{T}\left\vert
\frac{C_{M_{j}}(T)-C_{M_{j}}(0)}{C_{M_{j}}(0)}\right\vert .
\label{relative_conserve}
\end{equation}%
In this implementation, the time step is $\Delta T=0.02$, the mesh size is $%
\Delta X=600/2^{16}$, and the total running time is $10000$. The low-pass
filter is applied to all the variables $(\xi _{j},\overline{u}_{j})$ every $%
40$ time steps. Finally, our numerical results show that the maximum
relative errors in conservation of $(\xi _{1},\xi _{2},\overline{u}_{1},%
\overline{u}_{2})$ are $2.8$E$-7$, $1.9$E$-7$, $2.7$E$-7$, and $1.4$E$-7$,
respectively, that is, mass and momentum are conserved. After suppressing
the KH instability, an interfacial solitary wave solution can propagate for
a sufficiently long time [not shown here].

\section{KdV-Schr\"{o}dinger model }

\label{connect}

As mentioned in the conclusions and discussion \S\ \ref{sec:conclusion}, a particularly detailed prior effort of modeling  IW surface signature phenomena was undertaken in \citep{Craig2012JFM}. In this appendix, we compare the model of \citep{Craig2012JFM} with ours.
In \citep{Craig2004CRM,Craig2005CPAM,Craig2011NatHazd,Craig2012JFM},
by applying perturbation analysis and canonical transformations to a
Hamiltonian formulation of the two-layer Euler system,  a
resonantly-coupled system composed of a Korteweg--de Vries (KdV) equation
for the long-mode waves and a linear Schr\"{o}dinger equation (with an
internal soliton as its potential well) for the short-mode waves was constructed.
Narrow rough regions were found and interpreted as a result of the accumulation of
energy in the localized bound states of the Schr\"{o}dinger equation.

We also observe the accumulation of SW energy in our numerical
results as shown in figures \ref{Fig:OriginSetup_Amplitude} and \ref%
{Fig:ChoiSetup_Amplitude}. Thus, our numerical results are consistent with
the theory of energy accumulation for short SWs developed in \citep{Craig2012JFM}. This property of energy
accumulation is important for the generation of only one dominant wave
packet [figure \ref{Fig:ChoiSetup_Amplitude}], which may be responsible for
the narrow band of roughness generated by an IW as seen in
many field observations \citep{Gargettt1972JFM,Osborne1980Sci,Kropfli1999JGR}.

However, there exist two significant differences in the understanding of the surface
signature between our work and \citep{Craig2012JFM}. The first is the location of the surface ripples. In
\citep{Craig2012JFM}, due to the symmetry of the potential well (internal
soliton) in the linear Schr\"{o}dinger equation, the surface ripples are
located at the crest of the internal soliton. In contrast, our work shows
that due to the three types of asymmetric behavior of SWs \hyperref[asym1]{[1a]}-
\hyperref[asym3]{[1c]} (linear modulation theory), a sink of SW packets is
formed at the leading edge. These SWs can eventually accumulate at
the leading edge due to the class 3 triad resonance. The second is the dynamical
behavior of SWs in wavenumber $k$ space. In \citep{Craig2012JFM}, due to
the resonance condition (\ref{Eqn:resonce_cond}), the coupled KdV-Schr\"{o}%
dinger equation may only be valid for quasi-monochromatic surface waves
near the resonant $k_{\text{res}}$. However, we numerically show that, for a wide range of
wavenumbers, energy can eventually be concentrated near the resonant $k_{\text{res%
}}$ due to the modulation-resonance mechanism [see figure \ref%
{Fig:Mechanism_Diagram}(\textit{b})]. These resonant surface wave
packets can then propagate for a long time.


From the field observations, \cite{Osborne1980Sci} observed that, after the
passage of surface ripples, the sea surface becomes completely calm, like a
mill pond [the `mill pond' effect].
For this `mill pond' effect, \cite{Craig2012JFM} pointed out that the
passage of an internal soliton reflects the waves propagating in front of
it, effectively sweeping the surface of waves, and resulting in the observed
calmness of the sea after its passage. For the `mill pond'
effect, we present our hypothesis as follows: In the moving frame, the
small-amplitude SW packets with $k>k_{\text{res}}$ in front of the leading
edge [rightmost black arrows in figure \ref{Fig:Broad_near_field_theory}]
propagate towards the leading edge with a relatively small group velocity as
predicted by the linear modulation theory. Once colliding with the
large-amplitude dominant wave packet at the leading edge, these
high-wavenumber ($k>k_{\text{res}}$),\ small-amplitude SW packets may be
absorbed into the dominant wave packet. Therefore, in the rest frame,
after the passage of an internal soliton, waves with $k>k_{\text{res}}$ in front
of the leading edge may be absorbed into the dominant wave packet. Due to
very few waves with $k>k_{\text{res}}$\ traveling behind the dominant wave packet
(which is always located at the leading edge),
the `mill pond' effect may be formed after the passage of an
IW. However, due to the complicated set-up for the
boundary condition on the right-hand side of the computational domain
in front of the leading edge, it remains difficult to examine our hypothesis at
this stage.

Finally, dissipation may play an important role in the surface
signature. Since the typical wavelength of accompanying SWs is only about $%
0.1$-$1$ m for the underlying $1$ km internal soliton, these short SWs
should be dissipated quickly by the viscosity, within at most a few hours.
However, being generated by tidal flow, IWs usually can propagate thousands of
kilometers with the accompanying SWs. This, on the other hand, implies
that these accompanying SWs may accumulate energy  from the background environment
as pointed out in \citep{Craig2012JFM} and our work.

\bibliographystyle{jfm_first_modify_20151030}
\bibliography{Short_Format,myreference}

\begin{thebibliography}{46}
\expandafter\ifx\csname natexlab\endcsname\relax\def\natexlab#1{#1}\fi
\def\au#1{#1} \def\ed#1{#1} \def\yr#1{#1}\def\at#1{#1}\def\jt#1{\textit{#1}}
  \def\bt#1{#1}\def\bvol#1{\textbf{#1}} \def\vol#1{#1} \def\pg#1{#1}
  \def\publ#1{#1}\def\arxiv#1{#1}\def\org#1{#1}\def\st#1{\textit{#1}}

\bibitem[Alam(2012)]{Alam2012JFM}
{\sc \au{Alam, M.-R.}} \yr{2012}  \at{A new triad resonance between
  co-propagating surface and interfacial waves}.  \jt{J. Fluid Mech.}
  \bvol{691},  \pg{267--278}.

\bibitem[Alford {\em et~al.\/}(2015)Alford, Peacock, MacKinnon, Nash, Buijsman,
  Centuroni, Chao, Chang, Farmer, Fringer {\em et~al.\/}]{Alford2015Nature}
{\sc \au{Alford, M.~H.}, \au{Peacock, T.}, \au{MacKinnon, J.~A.}, \au{Nash,
  J.~D.}, \au{Buijsman, M.~C.}, \au{Centuroni, L.~R.}, \au{Chao, S.-Y.},
  \au{Chang, M.-H.}, \au{Farmer, D.~M.}, \au{Fringer, O.~B.} \& \au{others}}
  \yr{2015}  \at{The formation and fate of internal waves in the south china
  sea}.  \jt{Nature}  \bvol{521}~(7550),  \pg{65--69}.

\bibitem[Alpers(1985)]{Alpers1985Nature}
{\sc \au{Alpers, W.}} \yr{1985}  \at{Theory of radar imaging of internal
  waves}.  \jt{Nature}  \bvol{314}~(6008),  \pg{245--247}.

\bibitem[Apel {\em et~al.\/}(2007)Apel, Ostrovsky, Stepanyants \&
  Lynch]{Apel2007JASA}
{\sc \au{Apel, J.~R.}, \au{Ostrovsky, L.~A.}, \au{Stepanyants, Y.~A.} \&
  \au{Lynch, J.~F.}} \yr{2007}  \at{Internal solitons in the ocean and their
  effect on underwater sound}.  \jt{J. Acoust. Soc. Amer.}
  \bvol{121}~(695¨C722).

\bibitem[Bakhanov \& Ostrovsky(2002)]{Bakhanov2002JGR}
{\sc \au{Bakhanov, V.~V.} \& \au{Ostrovsky, L.~A.}} \yr{2002}  \at{Action of
  strong internal solitary waves on surface waves}.  \jt{J. Geophys. Res.}
  \bvol{107}~(3139).

\bibitem[Barros \& Choi(2009)]{Barros2009SAM}
{\sc \au{Barros, R.} \& \au{Choi, W.}} \yr{2009}  \at{Inhibiting shear
  instability induced by large amplitude internal solitary waves in two-layer
  flows with a free surface}.  \jt{Stud. Appl. Math}  \bvol{122}~(325-346).

\bibitem[Barros \& Gavrilyuk(2007)]{Barros2007SAM}
{\sc \au{Barros, R.} \& \au{Gavrilyuk, S.}} \yr{2007}  \at{Dispersive nonlinear
  waves in two-layer flows with free surface part ii. large amplitude solitary
  waves embedded into the continuous spectrum}.  \jt{Stud. Appl. Math}
  \bvol{119}~(213-251).

\bibitem[Benney(1977)]{Benney1977SAM}
{\sc \au{Benney, D.}} \yr{1977}  \at{A general theory for interactions between
  short and long waves}.  \jt{Stud. Appl. Math}  \bvol{56}~(1),  \pg{81--94}.

\bibitem[Caponi {\em et~al.\/}(1988)Caponi, Crawford, Yuen \&
  Saffman]{Caponi1988DTIC}
{\sc \au{Caponi, E.~A.}, \au{Crawford, D.~R.}, \au{Yuen, H.~C.} \& \au{Saffman,
  P.~G.}} \yr{1988}  \bt{Modulation of radar backscatter from the ocean by a
  variable surface current}. {\em Tech. Rep.\/}.  \org{DTIC Document}.

\bibitem[Chen(2005)]{Chen2005Efficient}
{\sc \au{Chen, T.}} \yr{2005}  \at{An efficient algorithm based on quadratic
  spline collocation and finite difference methods for parabolic partial
  differential equations}. PhD thesis, University of Toronto.

\bibitem[Choi {\em et~al.\/}(2009)Choi, Barros \& Jo]{Choi2009JFM}
{\sc \au{Choi, W.}, \au{Barros, R.} \& \au{Jo, T.-C.}} \yr{2009}  \at{A
  regularized model for strongly nonlinear internal solitary waves}.  \jt{J.
  Fluid Mech.}  \bvol{629}~(73-85).

\bibitem[Choi \& Camassa(1996)]{Choi1996JFM}
{\sc \au{Choi, W.} \& \au{Camassa, R.}} \yr{1996}  \at{Weakly nonlinear
  internal waves in a two-fluid system}.  \jt{J. Fluid Mech.}
  \bvol{313}~(83-103).

\bibitem[Choi \& Camassa(1999)]{Choi1999JFM}
{\sc \au{Choi, W.} \& \au{Camassa, R.}} \yr{1999}  \at{Fully nonlinear internal
  waves in a two-fluid system}.  \jt{J. Fluid Mech.}  \bvol{396}~(1-36).

\bibitem[Craig {\em et~al.\/}(2004)Craig, Guyenne \& Kalisch]{Craig2004CRM}
{\sc \au{Craig, W.}, \au{Guyenne, P.} \& \au{Kalisch, H.}} \yr{2004}  \at{A new
  model for large amplitude long internal waves}.  \jt{C. R. Mecanique}
  \bvol{332}~(525-530).

\bibitem[Craig {\em et~al.\/}(2005)Craig, Guyenne \& Kalisch]{Craig2005CPAM}
{\sc \au{Craig, W.}, \au{Guyenne, P.} \& \au{Kalisch, H.}} \yr{2005}
  \at{Hamiltonian long wave expansions for free surfaces and interfaces}.
  \jt{Commun. Pure Appl. Maths}  \bvol{58}~(1587-1641).

\bibitem[Craig {\em et~al.\/}(2011)Craig, Guyenne \& Sulem]{Craig2011NatHazd}
{\sc \au{Craig, W.}, \au{Guyenne, P.} \& \au{Sulem, C.}} \yr{2011}
  \at{Coupling between internal and surface waves}.  \jt{Nat. Hazards}
  \bvol{57}~(617-642).

\bibitem[Craig {\em et~al.\/}(2012)Craig, Guyenne \& Sulem]{Craig2012JFM}
{\sc \au{Craig, W.}, \au{Guyenne, P.} \& \au{Sulem, C.}} \yr{2012}  \at{The
  surface signature of internal waves}.  \jt{J. Fluid Mech.}
  \bvol{710}~(277-303).

\bibitem[Donato {\em et~al.\/}(1999)Donato, Peregrine \&
  Stocker]{Donato1999JFM}
{\sc \au{Donato, A.~N.}, \au{Peregrine, D.~H.} \& \au{Stocker, J.~R.}}
  \yr{1999}  \at{The focusing of surface waves by internal waves}.  \jt{J.
  Fluid Mech.}  \bvol{384},  \pg{27--58}.

\bibitem[Duda \& Farmer(1999)]{Duda1998DTIC}
{\sc \au{Duda, T.~F.} \& \au{Farmer, D.~M.}} \yr{1999}  \bt{The 1998
  {WHOI/IOS/ONR} {I}nternal {S}olitary {W}ave {W}orkshop: {C}ontributed
  {P}apers}. {\em Tech. Rep.\/}.  \org{DTIC Document}.

\bibitem[Duda {\em et~al.\/}(2004)Duda, Lynch, Irish, Beardsley, Ramp, Chiu,
  Tang \& Yang]{Duda2004IEEEJOE}
{\sc \au{Duda, T.~F.}, \au{Lynch, J.~F.}, \au{Irish, J.~D.}, \au{Beardsley,
  R.~C.}, \au{Ramp, S.~R.}, \au{Chiu, C.~S.}, \au{Tang, T.~Y.} \& \au{Yang,
  Y.~J.}} \yr{2004}  \at{Internal tide and nonlinear wave behavior in the
  continental slope in the northern south china sea}.  \jt{IEEE J. Ocean. Eng.}
   \bvol{29}~(1105-1131).

\bibitem[Dysthe {\em et~al.\/}(2008)Dysthe, Krogstad \&
  M{\"u}ller]{Dysthe2008oceanic}
{\sc \au{Dysthe, K.}, \au{Krogstad, H.~E.} \& \au{M{\"u}ller, P.}} \yr{2008}
  \at{Oceanic rogue waves}.  \jt{Annu. Rev. Fluid Mech.}  \bvol{40},
  \pg{287--310}.

\bibitem[Funakoshi \& Oikawa(1983)]{Funakoshi1983JPSJp}
{\sc \au{Funakoshi, M.} \& \au{Oikawa, M.}} \yr{1983}  \at{The resonant
  interaction between a long internal gravity wave and a surface gravity wave
  packet}.  \jt{J. Phys. Soc. Jpn}  \bvol{56}~(1982-1995).

\bibitem[Gargett \& Hughes(1972)]{Gargettt1972JFM}
{\sc \au{Gargett, A.~E.} \& \au{Hughes, B.~A.}} \yr{1972}  \at{On the
  interaction of surface and internal waves}.  \jt{J. Fluid Mech.}
  \bvol{52}~(01),  \pg{179--191}.

\bibitem[Gasparovic {\em et~al.\/}(1988)Gasparovic, Apel \&
  Kasischke]{Gasparovic1988JGR}
{\sc \au{Gasparovic, R.~F.}, \au{Apel, J.~R.} \& \au{Kasischke, E.~S.}}
  \yr{1988}  \at{An overview of the sar internal wave experiment}.  \jt{J.
  Geophys. Res.}  \bvol{93}~(12,304-12,316).

\bibitem[Guyenne(2006)]{Guyenne2006CRM}
{\sc \au{Guyenne, P.}} \yr{2006}  \at{Large-amplitude internal solitary waves
  in a two-fluid model}.  \jt{Comptes Rendus M{\'e}canique}  \bvol{334}~(6),
  \pg{341--346}.

\bibitem[Han \& Xu(2007)]{Han2007AMC}
{\sc \au{Han, H.} \& \au{Xu, Z.}} \yr{2007}  \at{Numerical solitons of
  generalized korteweg¨cde vries equations}.  \jt{Appl. Math. Comput.}
  \bvol{186}~(483-489).

\bibitem[Hashizume(1980)]{Hashizume1980JPSJp}
{\sc \au{Hashizume, Y.}} \yr{1980}  \at{Interaction between short surface waves
  and long internal waves}.  \jt{J. Phys. Soc. Jpn}  \bvol{48}~(631-638).

\bibitem[Hwung {\em et~al.\/}(2009)Hwung, Yang \& Shugan]{Hwung2009JFM}
{\sc \au{Hwung, H.-H.}, \au{Yang, R.-Y.} \& \au{Shugan, I.~V.}} \yr{2009}
  \at{Exposure of internal waves on the sea surface}.  \jt{J. Fluid Mech.}
  \bvol{626}~(1-20).

\bibitem[Jo \& Choi(2008)]{Jo2008SAM}
{\sc \au{Jo, T.-C.} \& \au{Choi, W.}} \yr{2008}  \at{On stabilizing the
  strongly nonlinear internal wave model}.  \jt{Stud. Appl. Math}
  \bvol{120}~(65-85).

\bibitem[Johnston {\em et~al.\/}(2003)Johnston, Merrifield \&
  Holloway]{Johnston2003JGR}
{\sc \au{Johnston, T.~S.}, \au{Merrifield, M.~A.} \& \au{Holloway, P.~E.}}
  \yr{2003}  \at{Internal tide scattering at the line islands ridge}.  \jt{J.
  Geophys. Res.}  \bvol{108}~(C11).

\bibitem[Kawahara {\em et~al.\/}(1975)Kawahara, Sugimoto \&
  Kakutani]{Kawahara1975JPSJ}
{\sc \au{Kawahara, T.}, \au{Sugimoto, N.} \& \au{Kakutani, T.}} \yr{1975}
  \at{Nonlinear interaction between short and long capillary-gravity waves}.
  \jt{J. Phys. Soc. Jpn.}  \bvol{39}~(5),  \pg{1379--1386}.

\bibitem[Kodaira {\em et~al.\/}(2016)Kodaira, Waseda, Miyata \&
  Choi]{Kodaira2016JFM}
{\sc \au{Kodaira, T.}, \au{Waseda, T.}, \au{Miyata, M.} \& \au{Choi, W.}}
  \yr{2016}  \at{Internal solitary waves in a two-fluid system with a free
  surface}.  \jt{J. Fluid Mech.}  \bvol{804},  \pg{201--223}.

\bibitem[Koop \& Butler(1981)]{Koop1981JFM}
{\sc \au{Koop, C.~G.} \& \au{Butler, G.}} \yr{1981}  \at{An investigation of
  internal solitary waves in a two-fluid system}.  \jt{J. Fluid Mech.}
  \bvol{112},  \pg{225--251}.

\bibitem[Kropfli {\em et~al.\/}(1999)Kropfli, Ostrovski, Stanton, Skirta, Keane
  \& Irisov]{Kropfli1999JGR}
{\sc \au{Kropfli, R.~A.}, \au{Ostrovski, L.~A.}, \au{Stanton, T.~P.},
  \au{Skirta, E.~A.}, \au{Keane, A.~N.} \& \au{Irisov, V.}} \yr{1999}
  \at{Relationships between strong internal waves in the coastal zone and their
  radar and radiometric signatures}.  \jt{J. Geophys. Res.}
  \bvol{104}~(3133-3148).

\bibitem[Lee {\em et~al.\/}(2007)Lee, Shugan \& An]{Lee2007JKrPS}
{\sc \au{Lee, K.-J.}, \au{Shugan, I.~V.} \& \au{An, J.-S.}} \yr{2007}  \at{On
  the interaction between surface and internal waves}.  \jt{J. Korean Phys.
  Soc.}  \bvol{51}~(616-622).

\bibitem[Lewis {\em et~al.\/}(1974)Lewis, Lake \& Ko]{Lewis1974JFM}
{\sc \au{Lewis, J.~E.}, \au{Lake, B.~M.} \& \au{Ko, D. R.~S.}} \yr{1974}
  \at{On the interaction of internal waves and surface gravity waves}.  \jt{J.
  Fluid Mech.}  \bvol{63}~(04),  \pg{773--800}.

\bibitem[Moore \& Lien(2007)]{Moore2007MMS}
{\sc \au{Moore, S.~E.} \& \au{Lien, R.-C.}} \yr{2007}  \at{Pilot whales follow
  internal solitary waves in the south china sea}.  \jt{Mar. Mamm. Sci.}
  \bvol{23}~(1),  \pg{193--196}.

\bibitem[M{\"u}ller {\em et~al.\/}(2005)M{\"u}ller, Garrett \&
  Osborne]{Muller2005Oceano}
{\sc \au{M{\"u}ller, P.}, \au{Garrett, C.} \& \au{Osborne, A.}} \yr{2005}
  \at{Rogue waves}.  \jt{Oceanography}  \bvol{18}~(3),  \pg{66}.

\bibitem[Osborne \& Burch(1980)]{Osborne1980Sci}
{\sc \au{Osborne, A.~R.} \& \au{Burch, T.~L.}} \yr{1980}  \at{Internal solitons
  in the andaman sea}.  \jt{Science}  \bvol{208}~(451-460).

\bibitem[Parau \& Dias(2001)]{Parau2001JFM}
{\sc \au{Parau, E.} \& \au{Dias, F.}} \yr{2001}  \at{Interfacial periodic waves
  of permanent form with free-surface boundary conditions}.  \jt{J. Fluid
  Mech.}  \bvol{437}~(325-336).

\bibitem[Perry \& Schimke(1965)]{Perry1965JGR}
{\sc \au{Perry, R.~B.} \& \au{Schimke, G.~R.}} \yr{1965}  \at{Large-amplitude
  internal waves observed off the northwest coast of sumatra}.  \jt{J. Geophys.
  Res.}  \bvol{70}~(10),  \pg{2319--2324}.

\bibitem[Phillips(1966)]{Phillips1966Dynamics}
{\sc \au{Phillips, O.~M.}} \yr{1966} {\em The Dynamics of the Upper Ocean\/}.
  \publ{Cambridge University Press}.

\bibitem[Phillips(1974)]{Phillips1974ARFM}
{\sc \au{Phillips, O.~M.}} \yr{1974}  \at{Nonlinear dispersive waves}.
  \jt{Annu. Rev. Fluid Mech.}  \bvol{6}~(93-110).

\bibitem[Sepulveda(1987)]{Sepulveda1987PhysFld}
{\sc \au{Sepulveda, N.}} \yr{1987}  \at{Solitary waves in the resonant
  phenomenon between a surface gravity wave packet and an internal gravity
  wave}.  \jt{Phys. Fluids}  \bvol{30}~(7).

\bibitem[Tanaka \& Wakayama(2015)]{Tanaka2015JFM}
{\sc \au{Tanaka, M.} \& \au{Wakayama, K.}} \yr{2015}  \at{A numerical study on
  the energy transfer from surface waves to interfacial waves in a two-layer
  fluid system}.  \jt{J. Fluid Mech.}  \bvol{763},  \pg{202--217}.

\bibitem[Whitham(1974)]{Whitham1974Linear}
{\sc \au{Whitham, G.~B.}} \yr{1974} {\em Linear and Nonlinear Waves\/}.
  \publ{A Wiley-Interscience Publication, New York}.

\end{thebibliography}

\end{document}